%% file: main.tex
\documentclass[a4paper,11pt]{article}
\pdfoutput=1
\usepackage{amssymb,amsmath,amsfonts,makeidx,placeins,pbox,multirow}
\usepackage{graphicx,rotate,subcaption,color,slashed,cite,caption,epstopdf,verbatim}
\usepackage{grffile}
\usepackage{longtable,tabu}
\usepackage{array}
\usepackage{pstricks}
\usepackage{color}
\usepackage{amsmath}
\usepackage{mathtools}
\usepackage{amssymb}
\usepackage{bbold}
\usepackage[normalem]{ulem}
\usepackage{caption}
\usepackage{subcaption}
\usepackage{dsfont}

\newcolumntype{P}[1]{>{\centering\arraybackslash}p{#1}}
\usepackage[colorlinks=true,
            linkcolor=magenta,
            urlcolor=blue,
            citecolor=blue]{hyperref}
\usepackage{cleveref}

\evensidemargin=\oddsidemargin
\newcommand{\beq}{\begin{equation}}
\newcommand{\eeq}{\end{equation}}
\newcommand{\bea}{\begin{eqnarray}}
\newcommand{\eea}{\end{eqnarray}}

\def\bar {\overline}

\def\to {\rightarrow}

\def\bea {\begin{eqnarray}}
\def\eea {\end{eqnarray}}
\def\n {\nonumber}

\def\st {\sin\theta}

\def\barr{\begin{array}}
\def\earr{\end{array}}
\def\to{\end{rightarrow}}

\def\gev{\ensuremath{\mathrm{Ge\kern -0.1em V}}}

\newcommand{\lfb}{\left(}
\newcommand{\rfb}{\right)}

\usepackage{color}
\definecolor{amber}{rgb}{1,0.45,0}

\usepackage{tikz}
\usetikzlibrary{positioning,arrows}
\usetikzlibrary{decorations.pathmorphing}
\usetikzlibrary{decorations.markings}

\begin{document}

\tikzset{
vector/.style={decorate, decoration={snake,amplitude=.6mm}, draw=red},
scalar/.style={dashed, draw=blue},
fermion/.style={draw=black, postaction={decorate},
        decoration={markings,mark=at position .55 with {\arrow[draw=black]{>}}}}
}
\begin{center}
{\Large \bf Exotic Decays and Collider Signatures of pNGB Scalars in the $SU(5)/SO(5)$ Composite Higgs Model\\
\vspace{0.3cm}
} 
\vspace*{0.4cm} {\sf $^{a}$Nilanjana Kumar\footnote{nilanjana.kumar@gmail.com}, $^{b}$Vandana Sahdev \footnote{vandanasahdev20@gmail.com}} \\
\vspace{6pt} {\small } {\em $^{a}$
Department of Physics, Chettinad Institute of Technology, Chettinad Academy of Research and Education, Tamilnadu 603103, India.


$^{b}$Department of Physics and Astrophysics, University of Delhi, Delhi-110007, India} \\
\normalsize
\end{center}
\vspace{-0.4cm}
\begin{abstract}
The nature of the Higgs boson, whether it is elementary or composite, will be investigated through precision measurements at the collider experiments. In composite Higgs scenarios, the Higgs may manifest as a pseudo Nambu-Goldstone boson (pNGB) arising from a strongly interacting sector. The $SU(5)/SO(5)$ Composite Higgs Model features a rich scalar sector, with the decay patterns of the scalars being heavily influenced by the way fermions are embedded in various representations of $SU(5)$. We discuss how masses 
of the pNGB scalars and their couplings depend functionally on the compositeness scale and the parameters of the strong sector. Unique decay modes of the scalars emerge from the model when the mixing among the various pNGB scalars 
is non-negligible. We present a comprehensive and thorough analysis of the fermiophilic and fermiophobic decay modes of the pNGB scalars. 
Significant differences are observed in the decay patterns of the two singly charged scalars. 
Further, the decay of one pNGB to another on-shell pNGB when masses exceed about $1$ TeV presents a rich phenomenology, leading to distinctive signatures at the colliders. 
In this context, 
the future muon collider offers a promising avenue for detecting pNGB scalars with masses larger than $1$ TeV, especially in final states involving $W/Z$ fatjets.

\end{abstract}
\vspace{-0.4cm}

\input{script}

\bibliographystyle{unsrt}
\bibliography{ref}

\end{document}

%% file: script.tex
\section{Introduction}
\label{sec:intro}

The last missing piece of the Standard Model (SM), the Higgs boson, was discovered by ATLAS \cite{Harrington:2013cca} and CMS \cite{CMS:2012qbp} at the Large Hadron Collider (LHC) in 2012 and, the properties of the SM have been observed with great precision at the LHC so far, 
with hints of disagreement in some measurements \cite{Muong-2:2021ojo,ATLAS:2022jtk,ParticleDataGroup:2024cfk}. Nevertheless, the question of whether the Higgs is a fundamental particle or composite persists. Moreover, in the Standard Model, the Higgs mass is not protected by any symmetry, thus, it receives large corrections at higher scales.
Also, the SM does not provide any dynamical explanation for the mechanism of electroweak symmetry breaking (EWSB). Composite Higgs Models \cite{Kaplan:1983fs,Dugan:1984hq,Contino:2010rs,Panico:2015jxa} offer an elegant solution to the hierarchy problem in Higgs physics by predicting a new strongly interacting sector
around the TeV energy scale. In these models, the Higgs emerges as a pseudo-Nambu-Goldstone boson of the strong sector \cite{Kaplan:1983fs}. The global symmetry is spontaneously broken, thereby, protecting the Higgs mass from corrections above the compositeness scale (\emph{f}). The angle of vacuum misalignment ($\theta$) is expressed as $\sin\theta=v/f$, where $v=246$ GeV, is the electroweak vacuum expectation value (v.e.v).

The minimal case, which is known as the Minimal Composite Higgs model, based on $SO(5)/SO(4)$ ~\cite{Agashe:2004rs,Panico:2011pw,Matsedonskyi:2012ym,Ferretti:2013kya} predicts only one pNGB scalar. 
Depending upon the global symmetry, the non-minimal scenarios predict new composite pNGB scalars. 
For example, in $SO(6)/SO(5)$~\cite{Banerjee:2017qod,Banerjee:2017wmg}, two pNGB scalars are obtained.
In other popular non-minimal Composite Higgs Models such as $SU(4)/ Sp(4)$ \cite{Bennett:2017kga}, $SU(5)/ SO(5)$\cite{Dugan:1984hq}, $SU(4)^{2}/SU(4)$ \cite{Ma:2015gra} etc., the breaking is introduced by the dynamics of the strong sector fermions \cite{Sanz:2017tco}.
These models not only offer a very rich scalar spectrum, they also predict many observations such as flavor 
changing neutral currents~\cite{Ma:2015gra}, dark matter~\cite{Cai:2020njb}, baryon asymmetry in the universe~\cite{Bruggisser:2018mus}, anomalous magnetic moment measurements~\cite{Doff:2015nru} etc.
Moreover, these scalars exhibit unique decay modes. Thus, the signatures of these pNGB scalars may or may not match with the 
standard searches of the exotic scalars performed by the LHC~\cite{ATLAS:2021jol,Cacciapaglia:2022bax}.

We choose to study the specific model $SU(5)/SO(5)$~\cite{Agugliaro:2018vsu,Ferretti:2014qta,Dugan:1984hq} which exhibits a very rich scalar sector. Here, in addition to a scalar doublet, the field content also includes one gauge singlet pseudo-scalar and three $SU(2)_L$ triplets.
At low energy, this model looks similar to the Georgi-Machacek (GM) Model~\cite{Englert:2013wga,Banerjee:2019gmr} 
with an additional singlet~\cite{Campbell:2016zbp}\footnote{However, there is a crucial difference between the scalars of the GM model and the $SU(5)/SO(5)$ scenario. The neutral component of the custodial triplet is CP-even, while the others are CP-odd, forbidding the singlet VEV and thus preventing tree-level coupling of new scalars to $W$ and $Z$ bosons in CP-conserving cases.}. 
The composite sector and the SM fermion sector are coupled via a linear mixing between the SM fermions and composite spin $1/2$ operators under the partial compositeness framework~\cite{Ferretti:2016upr}, which requires the existence of heavy fermionic resonances~\cite{Bizot:2018tds}.
Depending upon the embedding of the SM fermions in the representations of $SU(5)$, the scalars may or may 
not have couplings with the fermions. This has been extensively studied in Refs. \cite{Ferretti:2016upr,Agugliaro:2018vsu}. 
For example, if the left-handed and right-handed fermions are in the adjoint representation of $SU(5)$, 
the coupling of the pNGBs to the SM fermions vanishes, except with the SM Higgs. It is shown in Ref. \cite{Agugliaro:2018vsu} that in 
other possible representations, the couplings of the pNGB scalars to the SM fermions can be generated. Hence, depending upon the 
embedding of the fermions, the decays of the pNGB scalars are very distinct: either {\emph{fermiophilic}} or {\emph{fermiophobic}}.

Recently, Refs.~\cite{Banerjee:2022xmu,Cacciapaglia:2022bax} have studied a subset of the fermiophilic and fermiophobic scenarios in detail and discussed possible collider signatures and bounds from the LHC. 
The discussion in these papers is based on a simplified scenario where the value of the compositeness scale (\emph{f}) and the masses 
of the scalars (hence, the mass differences among the pNGB scalars) are fixed at certain values. However, the value of the compositeness scale is generally fixed by the allowed fine tuning (measured in terms of the parameter $\xi=v^2/f^2$) in the model, which depends heavily on the embedding of the fermions in different representations~\cite{Barnard:2017kbb}. Although the value of \emph{f} is constrained from the electroweak precision data, it is safe to take the range of the compositeness scale from 1 TeV to 5 TeV \cite{Agugliaro:2018vsu}.  
We explicitly calculate the masses of the pNGB scalars as functions of \emph{f} and $C_g$, where $C_g$ is the contribution 
of the gauge loops in the pNGB potential and it encodes the dynamics of the strong sector. The parameter 
$C_g$ is also influenced by the selected representation in the model being analyzed.
The effect of these two parameters appears in the branching ratios of the pNGB scalars that we study and some unique decay patterns of the scalars are identified. 
In the previous studies of the pNGB scalars, the effects of mixing between the gauge and mass eigenstates of the scalars have been neglected. 
We derive the mass eigenstates and the couplings of the pNGBs using the 
mixing angles in the charged and neutral sectors. These considerations lead to interesting decay modes of the pNGB scalars with masses of order TeV.

ATLAS and CMS measurements for the Higgs coupling constant put a limit on the coupling of the standard Higgs boson with the SM fermions and gauge bosons. For example, Ref.~\cite{Sanz:2017tco} combined the Run 1 and the recent Run 2 LHC data and found that different choices for fermion embedding lead to a spread of limits but a lower bound on the scale $f$ can be set to $600$ GeV. This limit, however, is obtained by combining different Composite Higgs Models. There is no dedicated search for pNGB scalars in the $SU(5)/SO(5)$ Composite Higgs Model. Nevertheless, ATLAS and CMS have placed limits on the charged and neutral scalars in other BSM models. The direct search limit on the charged Higgs comes from LEP, which is around 80 GeV \cite{ALEPH:2013htx}. Limits exist on the production cross-section of singly-charged scalars where they directly decay to fermions \cite{CMS:2019bfg,ATLAS:2012nhc,ATLAS:2018gfm,ATLAS:2018ntn,ATLAS:2015nkq,Ma:2015gra, ATLAS:2022pbd}. 
Amongst the searches for fermiophobically decaying charged scalars, the searches exist where the singly-charged scalars decay to $WZ$~\cite{ATLAS:2022zuc,CMS:2021wlt} and the doubly-charged scalars decay to $WW$~\cite{ATLAS:2023sua,CMS:2021wlt,ATLAS:2018ceg,ATLAS:2021jol}. These experimental limits can be applied to the pNGB scalars when they have similar decay patterns.
In all these scenarios, the limit on the scalars does not exceed $1$ TeV. The mass spectrum of the scalars that we study for specific collider signatures is above $1$ TeV. Hence, these limits are not applicable.

For a large value of \emph{f}, around $5$ TeV, the pNGB scalars have masses of order TeV. Hence, observation of these scalars at the LHC, at current c.m. energies, would be difficult due to low pair production cross section.
Although a high-energy $pp$ machine~\cite{Arkani-Hamed:2015vfh} may be a great option, the presence of abundant gauge bosons in the final state makes it challenging to observe multi-jet signals (with or without leptons) due to the large QCD background. 
Given these limitations, the muon collider~\cite{Delahaye:2019omf,Han:2022edd} emerges as a promising alternative for detecting pNGB scalars with unique signatures. The muon collider offers two main advantages. Unlike other lepton colliders that use electron beams, energy loss from synchrotron radiation is minimized in muon collider due to muon’s greater mass. This feature allows the collider to achieve high energy as well as high luminosity. Further, the muon collider also offers a clean environment for studying multi-jets. We discuss the pair production cross-sections of pNGB scalars at $3$ TeV and $6$ TeV centre-of-mass energies of the muon collider. The unique decay modes of the pNGBs offer signals rich with multiple $W/Z$ {\it fatjets}. 

\section{Composite Higgs Model $SU(5)/SO(5)$}
The non-minimal Composite Higgs Models predict additional pNGB scalars in addition to the standard Higgs doublet. 
The composite Higgs scenario based on the $SU(5)/SO(5)$ coset \cite{Agugliaro:2018vsu} leads to 14 pNGBs which 
decompose under the custodial $SU(2)_L \times SU(2)_R$ as
\begin{align}
14 \rightarrow (3,3)+ (2,2) +(1,1),
\end{align}
\emph{i.e.,} it includes a bi-triplet, the Higgs bi-doublet (H) and a singlet ($\eta$).
The bi-triplet, further, decomposes as 
\begin{align}
(3,3)\rightarrow 5+ 3 + 1, 
\end{align}
\emph{i.e.,} it includes a quintuplet $\eta_5\equiv (\eta^0_5, \eta^\pm_5, \eta^{\pm\pm}_5)$, 
a triplet $\eta_3\equiv (\eta^0_3, \eta^\pm_3)$ and a singlet $\eta_1^0$ of custodial $SU(2)_C$.

First, we consider a particular scenario of the $SU(5)/SO(5)$ Composite Higgs Model in which the left- and right-handed fermions 
are embedded in the adjoint representation of $SU(5)$. In this scenario, the minimal potential is generated at Leading Order (LO) and the coupling of the pNGB scalars (other than the SM Higgs) to SM fermions is forbidden.
This situation is often termed as a {\emph{fermiophobic}} scenario.
On the other hand, their coupling to SM gauge bosons results from their covariant derivative which can be expressed through the Lagrangian as
\begin{align}
\mathcal{L}_{\Phi \Phi V} &=\frac{ie}{s_W}W^{-\mu}\sum_{i,j}\left[\kappa^{\phi^0_i \phi^+_j}_{W}\phi^0_i\overleftrightarrow{\partial_\mu}\phi^+_j+\kappa^{\phi^-_i \phi^{++}_j}_{W}\phi^-_i\overleftrightarrow{\partial_\mu}\phi^{++}_j\right]+{\mathrm{h.c.}} \nonumber\\
& +\frac{ie}{s_Wc_W}Z^\mu \left[\sum_{i<j} \kappa^{\phi^0_i \phi^0_j}_{Z} \phi^0_i\overleftrightarrow{\partial_\mu} \phi^0_j+ \sum_{i,j} \left(\kappa^{ \phi^+_i \phi^{-}_j}_{Z} \phi^+_i\overleftrightarrow{\partial_\mu} \phi^{-}_j+\kappa^{ \phi^{++}_i \phi^{--}_j}_{Z} \phi^{++}_i\overleftrightarrow{\partial_\mu} \phi^{--}_j \right)\right]\nonumber\\
&-ieA^\mu \sum_{i}\left[ \phi^+_i\overleftrightarrow{\partial_\mu} \phi^-_i+2 \phi^{++}_i\overleftrightarrow{\partial_\mu} \phi^{--}_i\right].
\label{L_pipiV} 
\end{align}
In the above equation, $\phi^{0} = \{\eta, \eta_{1}^{0}, \eta_{3}^{0}, \eta_{5}^{0} \}, ~\phi^{\pm} = \{ \eta_{3}^{\pm}, \eta_{5}^{\pm} \}$ and $\phi^{\pm \pm} = \eta_{5}^{\pm \pm}$. 
In these models, we also encounter additional gauge bosons.
After integrating them out, we can derive couplings of the 
pNGB scalars to the Standard Model gauge bosons through a dimension-5 term such as
\begin{align}
\nonumber
\mathcal{L}_{ \Phi V\tilde{V}} &= \frac{e^2}{16 \pi^2 v} \bigg[ \phi ^0\left(\kappa^{ \phi^0}_{\gamma\gamma}F_{\mu\nu}\tilde{F}^{\mu\nu}+\frac{2}{s_Wc_W}\kappa^{ \phi^0}_{\gamma Z}F_{\mu\nu}\tilde{Z}^{\mu\nu}+\frac{1}{s_W^2c_W^2}\kappa^{ \phi^0}_{ZZ}Z_{\mu\nu}\tilde{Z}^{\mu\nu}\right.\nonumber\\
&\left.+\frac{2}{s_W^2}\kappa^{ \phi^0}_{WW}W^+_{\mu\nu}\tilde{W}^{-\mu\nu}\right)
+\sum_i \phi^+ \left( \frac{2}{s_W}\kappa^{ \phi^+}_{\gamma W}F_{\mu\nu}\tilde{W}^{-\mu\nu}+ \frac{2}{s_W^2c_W}\kappa^{ \phi^+}_{ZW}Z_{\mu\nu}\tilde{W}^{-\mu\nu}\right)+{\rm h.c.} \nonumber\\
&+ \frac{1}{s_W^2}\sum_i \phi^{++}\kappa^{ \phi^{++}}_{W^-W^-}W^-_{\mu\nu}\tilde{W}^{-\mu\nu}+{\rm h.c.}\bigg].
\label{vv1}
\end{align}
The Lagrangian involving coupling between the SM Higgs boson and EW gauge bosons can be expressed as,
\begin{align}
\mathcal{L}_{ hVV} =2M_W^2 \frac{h}{v}\left[\kappa^{h}_{WW} W_\mu^+W^{-\mu}+\kappa^{h}_{ZZ} \frac{1}{2c_W^2}Z_\mu Z^\mu\right]\,.  \label{L_piVV}
\end{align}
We refer to the complete {\emph{fermiophobic}} Lagrangian as, 
\begin{align}
\mathcal{L}_{F-phobic} &= \mathcal{L}_{\Phi \Phi V} + \mathcal{L}_{ \Phi V\tilde{V}}.
\end{align}
It is extensively discussed in Ref.~\cite{Agugliaro:2018vsu} that if the fermions are embedded in other representations (such as Adjoint(L)+ Symmetric (R)) of $SU(5)$, the scalar potential receives contributions from both the fermion and gauge loops. Hence, interaction Lagrangian in the \emph{fermiophilic} scenario generates couplings between the pNGBs and the SM fermions which can be expressed  as, 
\begin{align}
\mathcal{L}_{F-philic} &= \mathcal{L}_{\Phi \Phi V} + \mathcal{L}_{ \Phi V\tilde{V}} + \mathcal{L}_{\Phi f f }.
\end{align}
where, 
\begin{align}
\mathcal{L}_{\Phi f f } & =
\phi^0 \bigg[\bar{t}\left(\kappa^{ \Phi^0}_{t}+i \kappa^{ \Phi^0}_{t}\gamma_5\right)t+\bar{b}\left(\kappa^{ \Phi^0}_{b}+i\tilde{\kappa}^{ \Phi^0}_{b}\gamma_5\right)b\bigg] +
 \Phi^+\left[\kappa^{ \phi^+}_{\bar t b,L} \bar{t} P_L b + (L\leftrightarrow R)\right]+ {\rm h.c.}
\label{L_ffpi}
\end{align}
Here, couplings (denoted by $\kappa$) to only the third generation of the SM fermions are considered. 
Note that we assume there is no Majorana-type coupling to the leptons. Hence, pNGB scalar decays into leptons 
are forbidden. In the next section, we discuss the gauge and fermionic couplings of the pNGB scalars in detail. 
\section{Mass spectrum and Couplings of the pNGB Scalars}
In this section, we discuss the masses and mixing of the pNGBs in the scenario where both the left- and right-handed fermions are embedded in the adjoint representation of $SU(5)$. The mass matrices for the pNGBs are shown in Ref.~\cite{Agugliaro:2018vsu} and the masses are computed under the assumption that the mixing between the pNGBs in neutral and charged sectors is negligible. Here, we do not neglect the mixing and we derive the masses of the pNGB in physical basis and their couplings as a function of the mixing angles. 
\paragraph{\bf \underline{Mass of the Scalars}:}
The doubly-charged scalar remains unmixed and has mass\cite{Agugliaro:2018vsu},
\begin{align}
m_{\eta_5^{\pm \pm} } ^2  &= \frac{2}{3} m_h^2 \cot  ^2  2\theta + 4 \frac{C_g v^2 }{s^2_{\theta}}(3 g^2+ g'^2) + \frac{4}{3} \frac{C_g v^2}{s^2_{\theta}}  \left(3-c_{2\theta}\right) g'^2
\end{align}
The mixing between the two singly charged custodial eigenstates, $\eta^{\pm}_{3}$ and $\eta^{\pm}_{5}$, can be parameterized by the following mass matrix~\cite{Agugliaro:2018vsu},
\begin{equation}
    \mathcal{M}_{\pm}^{2} = 
    \begin{pmatrix}
        \frac{2}{3}m^{2}_{h}\cot^{2}{2\theta}+4C_{g}(3g^{2}+g'^{2}-\frac{g'^{2}}{3}c_{2\theta})\frac{v^{2}}{s^{2}_{\theta}} & 4C_g g'^{2}\frac{v^{2}}{s^{2}_{\theta}}\\
       4 C_g g'^{2}\frac{v^{2}}{s^{2}_{\theta}} &  \frac{2}{3}m^{2}_{h}\cot^{2}{2\theta}+4C_{g}(3g^{2}+\frac{g'^{2}}{2}+\frac{1}{6}g'^{2}c_{2\theta})\frac{v^{2}}{s^{2}_{\theta}} 
    \end{pmatrix}
\end{equation}
On diagonalisation, we obtain the mass eigenstates, which we denote as, $\eta^{\pm}_{1}$ and $\eta^{\pm}_{2}$. The rotation between mass basis and the gauge basis are defined by angle $k_{+}$, given by,
\begin{equation}
    \tan{2k_{+}} = \frac{2\mathcal{M}_{\pm 12}}{\mathcal{M}_{\pm 11}-\mathcal{M}_{\pm 22}},
\end{equation}
such that, 
\begin{align}
\eta^{\pm}_{1}= \cos{k_{+}} \eta^{\pm}_{3} - \sin{k_{+}} \eta^{\pm}_{5}~~~~~~
\eta^{\pm}_{2}= \sin{k_{+}} \eta^{\pm}_{3} + \cos{k_{+}}\eta^{\pm}_{5}. 
\end{align}
The masses, $m^{2}_{\eta^{\pm}_{1}}$ and $m^{2}_{\eta^{\pm}_{2}}$  are obtained, straightforwardly, from the mass matrix $\mathcal{M}_{\pm}^{2}$.

The neutral scalar, $\eta_3^0$, being CP even, does not mix with other scalars. In addition, the other singlet scalar, $\eta$, mixes with  $\eta^{0}_{1}$ and $\eta^{0}_{5}$ but the mixing is negligible. Hence, their masses are \cite{Agugliaro:2018vsu},
\begin{align}
m_{\eta_3^0}^2 &\approx  \frac{2}{3} m_h ^2 \cot^2 2\theta +4 \frac{C_g v^2}{\st^2} \left( (3g^2+ g'^2) + \frac{2}{3} g'^2 c_{2\theta} \right). \nonumber\\
m_\eta ^2  &\approx  \frac{5}{6} \frac{m_h^2}{s^2_{\theta}}-\frac{20}{3} C_g (3g^2+g'^2) \frac{v^2}{s^2_{\theta}}. \nonumber\\
\end{align}
On the other hand, there exists a significant mixing between $\eta^{0}_{1}$ and $\eta^{0}_{5}$. The $2\times2$ mass matrix is~\cite{Agugliaro:2018vsu},
\begin{equation}
    \mathcal{M}^{2} = 
    \begin{pmatrix}
        \frac{1}{6}\frac{m^{2}_{h}}{s^{2}_{\theta}}+4C_{g}(3g^{2}+g'^{2})\frac{v^{2}}{s^{2}_{\theta}} & \frac{8}{3}\sqrt{2} C_g g'^{2}\frac{v^{2}}{s^{2}_{\theta}}\\
        \frac{8}{3}\sqrt{2} C_g g'^{2}\frac{v^{2}}{s^{2}_{\theta}} &  \frac{1}{6}\frac{m^{2}_{h}}{s^{2}_{\theta}}+4C_{g}(3g^{2}+g'^{2})\frac{v^{2}}{s^{2}_{\theta}} -\frac{8}{3} C_g g'^{2} \frac{v^{2}}{s^{2}_{\theta}}  
    \end{pmatrix}.
\end{equation}
The mass eigenstates $\lfb \eta_{1}, \eta_{2}\rfb$ are related to the gauge eigenstates $\lfb \eta^{0}_{1}, \eta^{0}_{5}\rfb$ via a rotation by 
an angle, $k_{0}$, given by,
\begin{equation}
    \tan{2k_{0}} = \frac{2\mathcal{M}_{12}}{\mathcal{M}_{11}-\mathcal{M}_{22}},
\end{equation}
and the mass eigenstates are
\begin{align}
\eta_{1}= \cos{k_{0}}~\eta^{0}_{1} - \sin{k_{0}}~\eta^{0}_{5},~~~
\eta_{2}= \sin{k_{0}}~\eta^{0}_{1} + \cos{k_{0}}~\eta^{0}_{5}.
\end{align}
It is straightforward to obtain the masses, $m^{2}_{\eta_{1}}$ and $m^{2}_{\eta_{2}}$, from the mass matrix, $\mathcal{M}^{2}$.
\paragraph{\bf \underline {Coupling of the Scalars}:}
Having obtained the mass spectrum, we derive the couplings of the pNGB scalars in terms of the model parameters and express them in the mass basis. 
The fermiophobic couplings of the pNGB scalars are listed in Table \ref{tab:coup_scalar1} and Table \ref{tab:coup_scalar2}.
Note that the mixing angles in the charged and the neutral sector depend on the parameters $C_g$ and \emph{f}, as $\sin\theta\sim v/f$. 
The couplings of the scalars are also function of the mixing angle $k_0$ and $k_{+}$ in the mass basis. In the fermiophilic scenario, the effective potential receives contributions from heavy fermions running in the loop. Consequently, the scalar masses and couplings acquire additional corrections. Nevertheless, because of the large masses of these fermions, their impact on the pNGB scalar masses remains negligible.. However, in the limit of small mixing angles, that is small
$k_{+}$ and $k_0$, we get back the expressions where the mass eigenstates are assumed to be the same as the gauge eigenstates.
\begin{table}[h]
    \centering
    \begin{tabular}{c c c c }
    \hline
        & \multicolumn{2}{c}{$\kappa_{W}^{\phi^{0}_{i}\phi^{+}_{j}}$} & \multicolumn{1}{c}{$\kappa_{W}^{\phi^{-}_{i}\phi^{++}_{j}}$} \\
        \hline
        & $\eta_{1}^{+}$ &$\eta_{2}^{+}$ &$\eta_{5}^{++}$\\
        \hline
        $\eta_{3}^{0}$ & $\frac{-i}{2}c_{k_{+}} - \frac{c_{\theta}}{2}s_{k_{+}}$ & $\frac{-i}{2}s_{k_{+}} + \frac{c_{\theta}}{2}c_{k_{+}}$  &\\
        $\eta_{1}$ & $\frac{c_{\theta}}{2\sqrt{3}}s_{k_{0}}c_{k_{+}}+ \frac{i\sqrt{3}}{2} s_{k_{0}}s_{k_{+}} + \sqrt{\frac{2}{3}}c_{\theta}c_{k_{0}}c_{k_{+}}$ &  $\frac{c_{\theta}}{2\sqrt{3}}s_{k_{0}}s_{k_{+}}- \frac{i\sqrt{3}}{2} s_{k_{0}}c_{k_{+}} + \sqrt{\frac{2}{3}}c_{\theta}c_{k_{0}}s_{k_{+}}$ &\\
        $\eta_{2}$ &  $\frac{-c_{\theta}}{2\sqrt{3}}c_{k_{0}}c_{k_{+}}- \frac{i\sqrt{3}}{2} c_{k_{0}}s_{k_{+}} + \sqrt{\frac{2}{3}}c_{\theta}s_{k_{0}}c_{k_{+}}$&   $\frac{-c_{\theta}}{2\sqrt{3}}c_{k_{0}}s_{k_{+}}+ \frac{i\sqrt{3}}{2} c_{k_{0}}c_{k_{+}} + \sqrt{\frac{2}{3}}c_{\theta}s_{k_{0}}s_{k_{+}}$&\\
        $\eta$ &  &  &\\
        $\eta_{1}^{+}$ & & &$\frac{c_{\theta}}{\sqrt{2}}c_{k_{+}} + \frac{i}{\sqrt{2}}s_{k_{+}}$\\
         $\eta_{2}^{+}$ & & &$\frac{c_{\theta}}{\sqrt{2}}s_{k_{+}} - \frac{i}{\sqrt{2}}c_{k_{+}}$\\
        \hline
    \end{tabular}  
\end{table}
\begin{table}[h]
    \centering
    \begin{tabular}{c c c c c c c c c }
    \hline
         & \multicolumn{2}{c}{$\kappa_{Z}^{\phi^{0}_{i}\phi^{+}_{j}}$} & \multicolumn{2}{c}{$\kappa_{Z}^{\phi^{+}_{i}\phi^{-}_{j}}$} & $\kappa_{Z}^{\phi^{++}_{i}\phi^{--}_{j}}$ \\
         \hline
         & $\eta_{1}$ & $\eta_{2}$  & $\eta_{1}^{-}$ & $\eta_{2}^{-}$ & $\eta_{5}^{--}$\\
         \hline
         $\eta_{3}^{0}$ & $\frac{-i c_{\theta}}{\sqrt{3}}s_{k_{0}}+i \sqrt{\frac{2}{3}} c_{\theta}c_{k_{0}}$ & $\frac{i c_{\theta}}{\sqrt{3}}c_{k_{0}}+i \sqrt{\frac{2}{3}} c_{\theta}s_{k_{0}}$&&&\\
         $\eta_{1}^{+}$ &&& $\frac{-c_{2W}}{2} $& $\frac{-ic_{\theta}}{2}$&\\
         $\eta_{2}^{+}$ &&&$\frac{ic_{\theta}}{2}$& $\frac{c_{2W}}{2}(c^{2}_{k_{+}}-s^{2}_{k_{+}})$&\\
         $\eta_{5}^{++}$ &&&&&$-c_{2W}$\\
         \hline
    \end{tabular}
    \caption{$\Phi \Phi V$ couplings in the mass basis of the pNGB scalars. }
    \label{tab:coup_scalar1}
\end{table}
\begin{table}[h]
    \centering
    \begin{tabular}{c c c c c }
    \hline
         & $\kappa^{ \phi^0}_{\gamma\gamma}$ & $\kappa^{ \phi^0}_{\gamma Z}$ & $\kappa^{ \phi^0}_{Z Z}$\\
    \hline
         $\eta_{1}$ & $\frac{2 s_{\theta}}{\sqrt{3}}s_{k_{0}}+\sqrt{\frac{2}{3}}s_{\theta}c_{k_{0}}$ & $\frac{c_{2W} s_{\theta}}{\sqrt{3}}s_{k_{0}}+\frac{c_{2W} s_{\theta}}{\sqrt{6}}c_{k_{0}}$ & $\frac{-(1-3c_{4W} + 2 c_{2 \theta}) s_{\theta}}{12\sqrt{3}}s_{k_{0}} + \frac{(1+6c_{4W} - 7 c_{2 \theta}) s_{\theta}}{24\sqrt{6}}c_{k_{0}}$ \\  
         $\eta_{2}$ & $\frac{-2 s_{\theta}}{\sqrt{3}}c_{k_{0}}+\sqrt{\frac{2}{3}}s_{\theta}s_{k_{0}}$& $\frac{-c_{2W} s_{\theta}}{\sqrt{3}}c_{k_{0}}+\frac{c_{2W} s_{\theta}}{\sqrt{6}}s_{k_{0}}$ & $\frac{(1-3c_{4W} + 2 c_{2 \theta}) s_{\theta}}{12\sqrt{3}}c_{k_{0}} + \frac{(1+6c_{4W} - 7 c_{2 \theta}) s_{\theta}}{24\sqrt{6}}s_{k_{0}}$ \\
         $\eta$ &  $\sqrt{\frac{2}{5}}s_{\theta}$ & $-\frac{c_{2W} s_{\theta}}{\sqrt{10}}$ & $\frac{(c_{4W} + 3 c^{2}_{ \theta}) s_{\theta}}{4\sqrt{10}}$\\
         \hline
    \end{tabular}
\vspace{0.2cm}
    \begin{tabular}{c c c c c}
         &$\kappa^{ \phi^0}_{W+W-}$ & $\kappa^{ \phi^+}_{\gamma W-}$ & $\kappa^{ \phi^+}_{Z W-}$& $\kappa^{ \phi^{++}}_{W- W-}$\\
    \hline
         $\eta_{1}$ & $\frac{ s^{3}_{\theta}}{6\sqrt{3}}s_{k_{0}}-\frac{ 7s^{3}_{\theta}}{12\sqrt{6}}c_{k_{0}}$ &&&\\  
         $\eta_{2}$ & $\frac{- s^{3}_{\theta}}{6\sqrt{3}}c_{k_{0}}-\frac{ 7s^{3}_{\theta}}{12\sqrt{6}}s_{k_{0}}$ &&&\\
         $\eta$ & $-\frac{(3c_{2\theta}+5) s^{2}_{\theta}}{8\sqrt{10}}$ &&&\\
         $\eta_{1}^{+}$ & &-$\frac{s_{2\theta}}{4}c_{k_{+}}-\frac{i s_{2\theta}}{2}s_{k_{+}}$ &$\frac{s^{2}_{W}s_{2\theta}}{4}c_{k_{+}}-\frac{i(c_{2W}-c_{2\theta} - 2)s_{\theta}}{12}s_{k_{+}}$ &\\
         $\eta_{2}^{+}$ & &-$\frac{s_{2\theta}}{4}s_{k_{+}}+\frac{i s_{2\theta}}{2}c_{k_{+}}$ &$\frac{s^{2}_{W}s_{2\theta}}{4}s_{k_{+}}+\frac{i(c_{2W}-c_{2\theta} - 2)s_{\theta}}{12}c_{k_{+}}$ &\\
         $\eta_{5}^{++}$ & & & & -$\frac{s^{3}_{\theta}}{3\sqrt{2}}$\\
         \hline
    \end{tabular}
    \caption{$\Phi V\tilde V$ couplings in the mass basis of the pNGB scalars.}
    \label{tab:coup_scalar2}
\end{table}

In the fermiophilic scenario, the effective potential receives contributions from heavy fermions running in the loop. Consequently, the scalar mass spectrum changes. 
However, Ref.\cite{Agugliaro:2018vsu} demonstrates that for specific values of parameters in the LO potential, the mass spectrum remains identical to that in the fermiophobic case, where left-handed and right-handed fermions are embedded in the adjoint representation of 
$SU(5)$. The couplings of the pNGB scalars to the third generation SM quarks are parameterized as,
\begin{align}
k^{\phi^0}_t=c_t \frac{m_t}{f},~~~ k^{\phi^0}_b=c_b \frac{m_b}{f}, ~~~k^{\phi^+}_{\bar t b}=c_{tb} \frac{m_t}{f},
\end{align}
at the leading order. The exact values of $c_t$, $c_b$ and $c_{tb}$ are obtained after integrating out the heavy fermions in the loop and that depends on the details of the embedding. We want to focus on the highest reachable branching ratio of the pNGBs to SM fermions, in the fermiophilic scenario and hence, we chose $c_i=1$, for all $i$.
We list the fermiophilic couplings in Table~\ref{table:coup_fermion} and show their approximate values at $f=2.5$ TeV and $5$ TeV in the mass basis. 

\begin{table}[h!]
\centering
\begin{tabular}{|c|c|c|}
\hline
{\bf Coupling} & {\bf Value at $f=2.5$ TeV} & {\bf Value at $f=5$ TeV} \\
\hline
$k^{\eta^{+}_{2}}_{\bar{t}b}$ = $c_{tb}(c_{k_{+}}+s_{k_{+}}) \frac{m_t}{f}$ & 0.098 & 0.049\\
$k^{\eta^{+}_{1}}_{\bar{t}b}$ = $c_{tb}(c_{k_{+}}-s_{k_{+}}) \frac{m_t}{f}$ & 0.00017 & 0.000083\\
$k^{\eta_{2}}_{t}$ = $c_t(c_{k_{0}}+s_{k_{0}}) \frac{m_t}{f}$ & 0.096 & 0.048\\
$k^{\eta_{1}}_{t}$ = $c_t(c_{k_{0}}-s_{k_{0}}) \frac{m_t}{f}$ & 0.016 & 0.008\\
$k^{\eta_{2}}_{b}$ = $c_b(c_{k_{0}}+s_{k_{0}}) \frac{m_b}{f}$ & 0.002 & 0.001\\
$k^{\eta_{1}}_{b}$ = $c_b(c_{k_{0}}-s_{k_{0}}) \frac{m_b}{f}$ & 0.0004 & 0.0002\\

$k^{\eta^{0}_{3}}_{b}$,$k^{\eta}_{b}$ = $c_b\frac{m_b}{f}$ & 0.0019 & 0.001\\
$k^{\eta^{0}_{3}}_{t}$, $k^{\eta}_{t}$ = $c_t\frac{m_t}{f}$ & 0.069 & 0.034\\
\hline
\end{tabular}
\caption{Fermionic ($\Phi f \bar{f}$) couplings of the pNGB scalars expressed in the mass basis at two different values of \emph{f} and for $c_t=c_b=c_{tb}=1$. The couplings for the conjugate fields have the same values.}
\label{table:coup_fermion}
\end{table}
It is important to observe that the fermiophilic couplings diminish in strength when \emph{f} is large, as evident from Table \ref{table:coup_fermion}.
We also observe that as a result of the mixing, the coupling of $\eta^{+}_{1}$ to $\bar{t}b$ is exceptionally suppressed, 
in contrast to the coupling of $\eta^{+}_{2}$ to $\bar{t}b$. 
In the neutral sector, the couplings of $\eta_{1}$ and $\eta_{2}$ with the SM quarks differ by only 
one order of magnitude. Hence, in the fermiophilic scenario, the decay modes of $\eta^{\pm}_{1}$ and $\eta^{\pm}_{2}$ are expected to be different whereas the decay modes of $\eta_{1}$ and $\eta_{2}$ should not differ much.
Note that, the total decay width of the scalars in the fermiophilic case will be function of both $c_i ~(i=t,b,tb)$ and the parameter, $C_g$ along with other model parameters, mixing angles $k_{+}$ and $k_0$, and $f$. Whereas, in the fermiophobic scenario, decay width depends on  the gauge contribution $C_g$ only along with $k_{+}$ and $k_0$, and $f$. We discuss the decays of the pNGB in the fermiophilic and fermiophobic scenario in the next section.
\section {Decays of the pNGB Scalars}
In the previous section, we demonstrated how the masses, mixing and couplings of the pNGB scalars are a function 
of the model parameters, $C_g$ and \emph{f}, where \emph{f} is present through $\sin{\theta} = v/f$. 
The value of these model parameters leads to specific patterns of mixing, parameterized by $k_{+}$ and $k_{0}$, 
and the mass splittings among the pNGB scalars which crucially affects the decay modes 
of these scalars. Ref. \cite{Agugliaro:2018vsu} presented a comprehensive analysis of the mass differences 
among these pNGB scalars and explored how these differences are influenced by the 
value of $C_g$ across all the possible representations.

\begin{figure}[h!]
\begin{center}
\includegraphics[height=4.5cm,width=5.5cm]{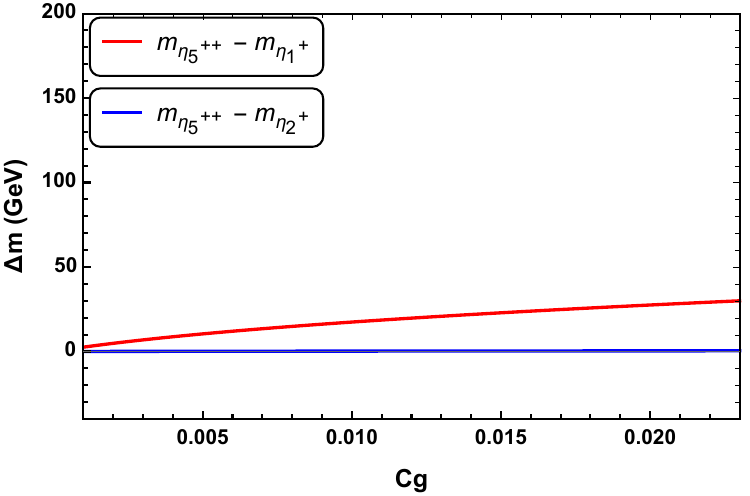}
\includegraphics[height=4.5cm,width=5.5cm]{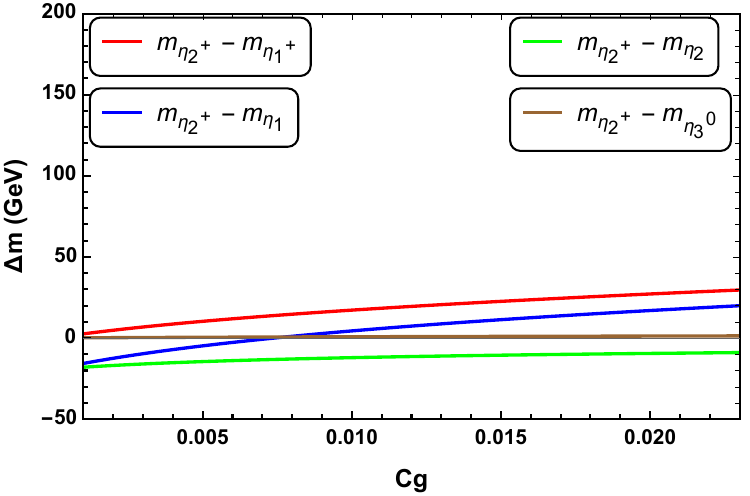}
\includegraphics[height=4.5cm,width=5.5cm]{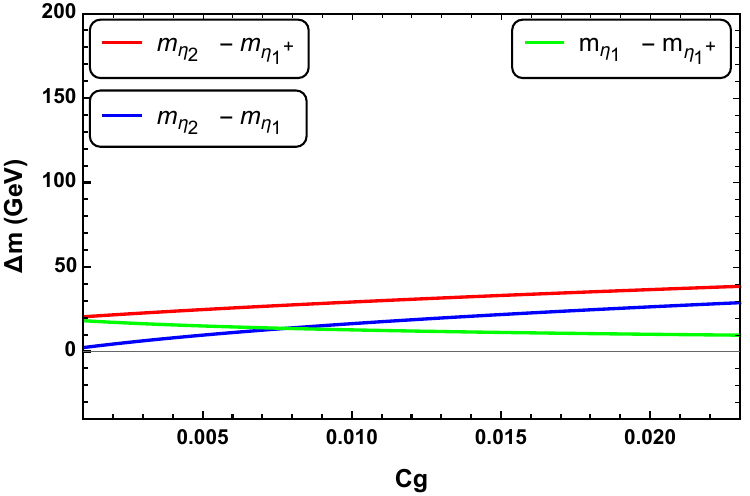}\\
\includegraphics[height=4.5cm,width=5.5cm]{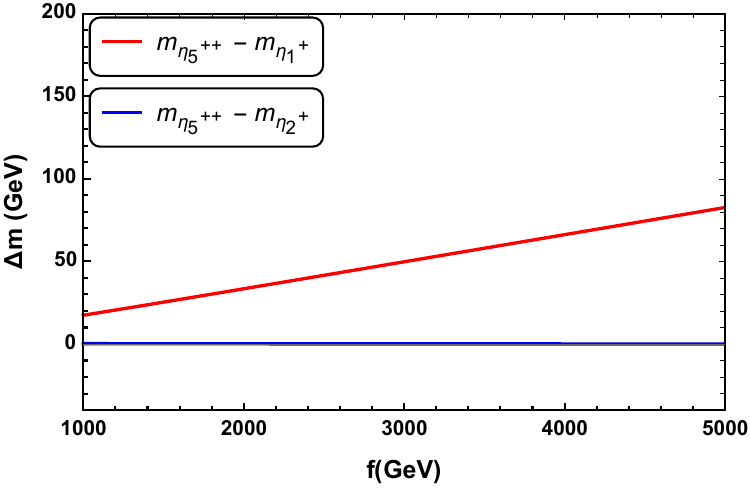}
\includegraphics[height=4.5cm,width=5.5cm]{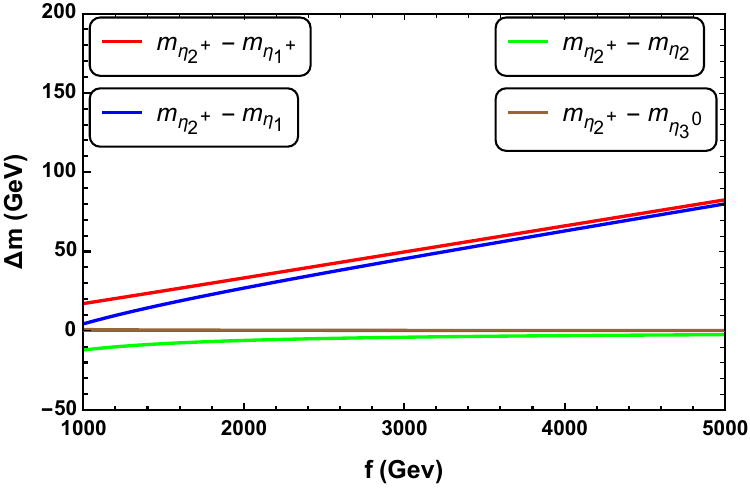}
\includegraphics[height=4.5cm,width=5.5cm]{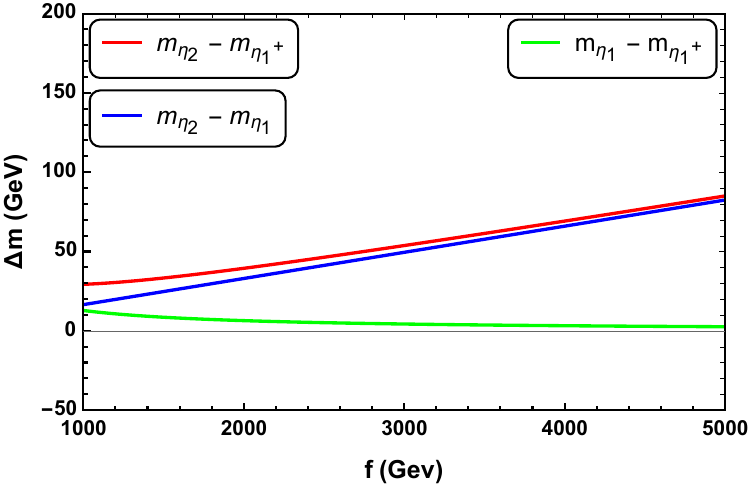}
\end{center}    
    \caption{Mass difference among the scalar pNGBs as a function of $C_g$ with $f=1$ TeV (top panel) and as a function 
    of \emph{f} with $C_g=0.01$ (bottom panel).}
    \label{fig:massdiff}
\end{figure}

In Table \ref{table:masses}, we show the mass spectrum at different benchmark values of the parameters, $C_g$ and \emph{f}.
\begin{table}[h!]
\centering
\begin{tabular}{|c|c|c|c|c|c|c|}
\hline
Mass (GeV) &\multicolumn{2}{c|}{$f=1$ TeV} & \multicolumn{2}{c|}{$f=2.5$ TeV} &\multicolumn{2}{c|}{$f=5$ TeV} \\
\hline
-&$C_g=0.01$ &$C_g=0.02$&$C_g=0.01$&$C_g=0.02$&$C_g=0.01$&$C_g=0.02$\\
\hline
$\eta_1^{\pm}$ & 286.3 & 358.7 & 743.0 & 918.4 & 1493.8 & 1843.1 \\
$\eta_2^{\pm}$ & 310.6 & 397.1 & 801.7 & 1012.3 & 1610.6 & 2030.4\\
$\eta_5^{\pm \pm}$ & 307.4 & 392.1 & 793.5 & 999.3 & 1594.1 & 2004.1\\
$\eta$ & 349.5 & 171.8 & 873.7 & 429.5 & 1747.5 & 859.0 \\
$\eta_1$ & 302.8 & 374.8 & 757.0 & 936.9 & 1514.1 & 1873.8 \\
$\eta_2$ & 319.3 & 401.2 & 798.2 & 1003 & 1596.5 & 2005.9 \\
$\eta_3^0$ & 306.4 & 390.5 & 793.1 & 998.7 & 1593.9 & 2003.8 \\
\hline
\end{tabular}
\caption{Mass of the pNGB scalars in GeV at different benchmark scenarios.}
\label{table:masses}
\end{table}
In Fig.~\ref{fig:massdiff}, we plot the mass differences between the scalars as a function of $C_g$ (top panel), 
and, as a function of \emph{f} (bottom panel).
It is clear from both, Table \ref{table:masses} and Fig.~\ref{fig:massdiff}, 
that the masses and the mass differences are highly sensitive to the parameters, $C_g$ and \emph{f}, particularly \emph{f}.
Thus, while analysing the decay modes, fixing the masses at one certain value or assuming a fixed mass splitting, 
is not always correct. A more insightful approach for analyzing the decays of the scalars is to include the effect of various mixing and mass-splittings, 
which we discuss in the next subsection.

Before proceeding further, we reflect upon a few nuances.  
The singlet scalar, $\eta$, receives a negative mass correction from the gauge sector \cite{Agugliaro:2018vsu}. 
Hence, as shown in Table \ref{table:masses}, as the value of $C_g$ increases, the mass of $\eta$ decreases.
At a particular value of $C_g$, the singlet becomes tachyonic and this places an upper limit on $C_g$ \emph{viz.,} $C_g<0.023$.
Hence, while observing the decay modes of the pNGB scalars, we vary $C_g$ between $0.001-0.023$. We also observe from Table \ref{table:masses} that 
the lowest mass scalar is either $\eta$ or $\eta_1$. We, now, describe the decay modes of the pNGB scalars, in detail.
\subsection{Branching Ratios of $\eta_{5}^{\pm\pm}$}
\begin{figure}[h!]
\includegraphics[height=5.5cm,width=5.7cm]{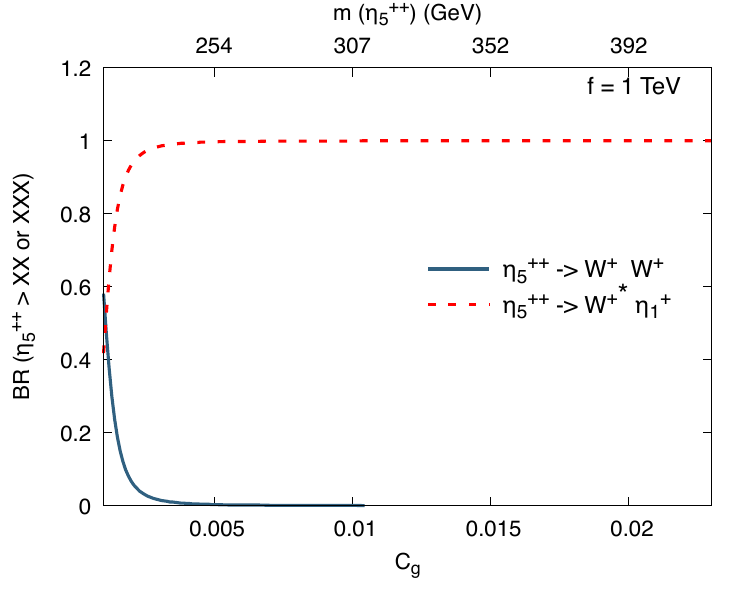}
\includegraphics[height=5.5cm,width=5.7cm]{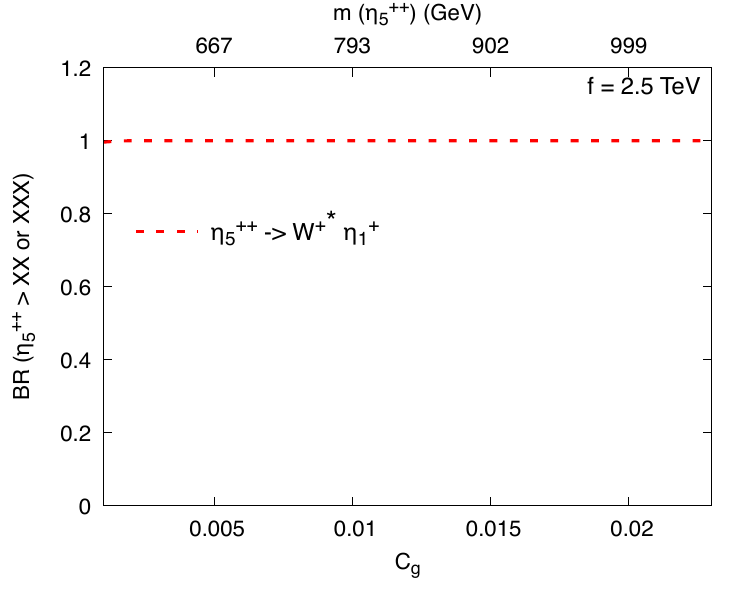}
\includegraphics[height=5.5cm,width=5.7cm]{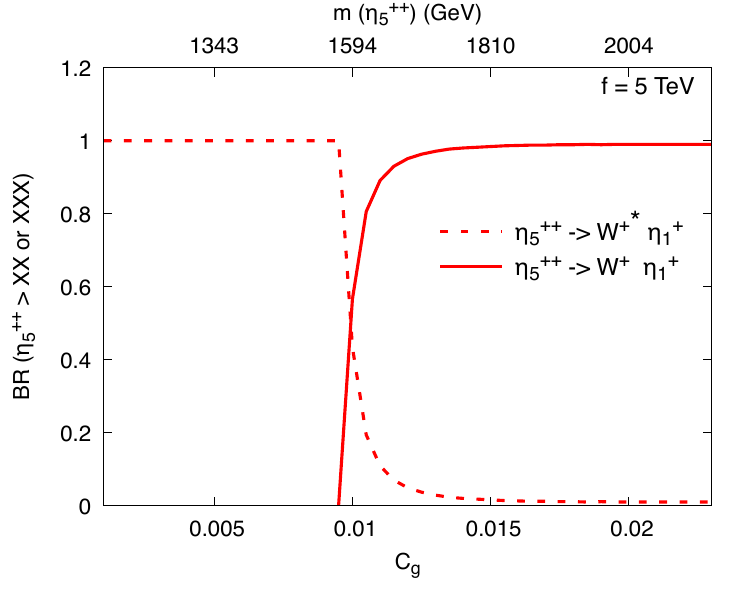}\\
\includegraphics[height=5.5cm,width=5.7cm]{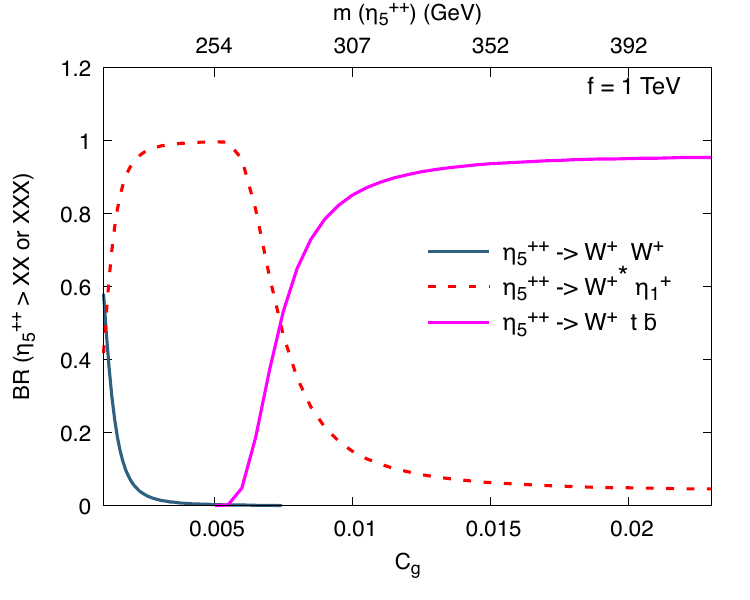}
\includegraphics[height=5.5cm,width=5.7cm]{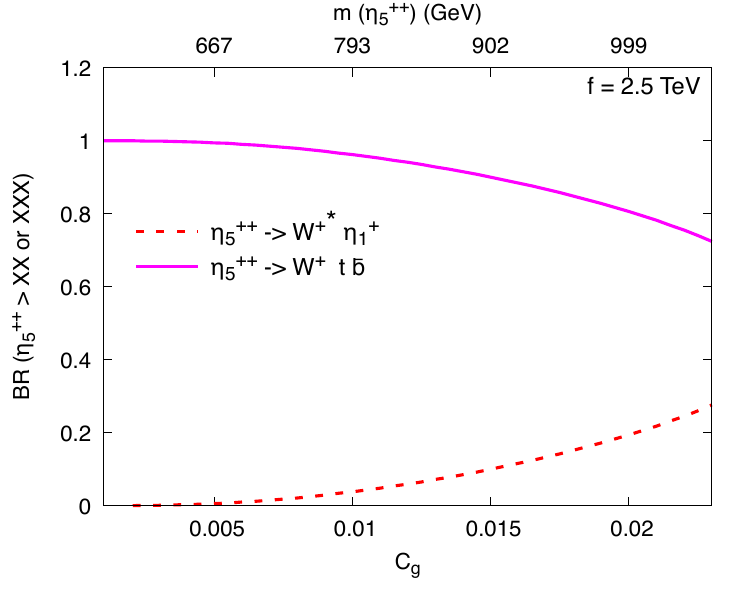}
\includegraphics[height=5.5cm,width=5.7cm]{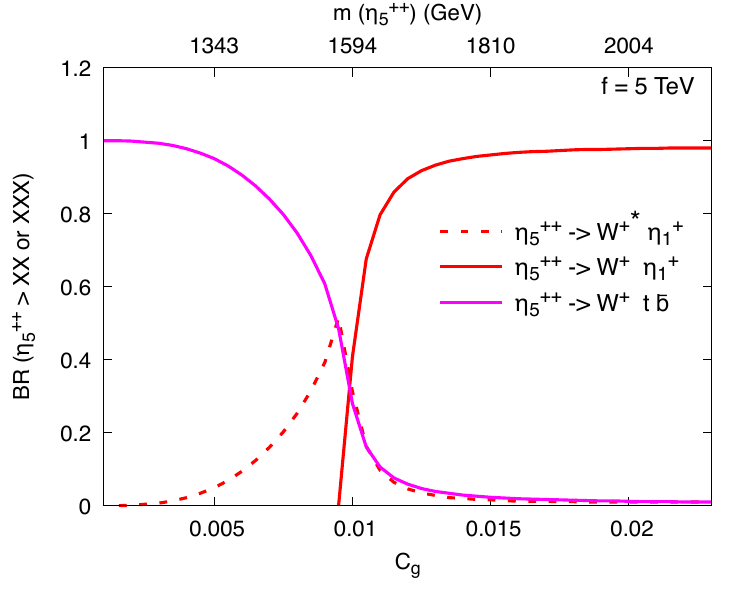}
    \caption{ Branching ratios of $\eta_{5}^{\pm\pm}$ for the {\it fermiophobic} scenario (top panel) and the \emph{fermiophilic} scenario (bottom panel) 
    at \emph{f} : 1 TeV (left), 2.5 TeV (middle) and 5 TeV (right).}
    \label{fig:n5pp}
    \end{figure}

The branching ratios of the doubly charged scalar $\eta_{5}^{\pm\pm}$ are plotted in Fig.~\ref{fig:n5pp}, 
for the fermiophobic scenario (top panel) and the fermiophilic scenario (bottom panel)
. The branching ratios are varied 
as a function of the pNGB mass and its various couplings, both affected by the parameters $C_g$ and $f$. 
Even if $f$ is fixed, the mass of the pNGB varies with  $C_g$, as depicted by the auxiliary X-axes in these plots. We have fixed \emph{f} 
at three different values: $1$ TeV (left), $2.5$ TeV (middle) and $5$ TeV (right) which covers the range of pNGB mass $\sim 200-2000$ GeV.
We discuss the fermiophobic scenario first (Fig.~\ref{fig:n5pp}, top panel). 
In the absence of the fermiophilic couplings, $\eta_5^{\pm\pm}$ decays to two gauge bosons or a pNGB scalar and an off-shell gauge boson, 
the latter dominating whenever sufficient phase space is available.
From the mass-differences depicted in Fig.~\ref{fig:massdiff}, we see that the relation $m_{\eta_{5}^{\pm \pm} }>m_{\eta_1^{\pm} }$ holds true for 
any value of $C_g$ and \emph{f}. 
On the other hand, the mass-splitting between $m_{\eta_{5}^{\pm\pm}}$ and $m_{\eta_2^{\pm}}$ is almost negligible.
Hence, the doubly charged scalar preferentially decays to $\eta_1^{\pm}$ and an off-shell $W$ boson leading to the following three-body decays,
at $f=1$ TeV and $2.5$ TeV:
\bea
\eta_{5}^{\pm \pm}~ &\rightarrow & W^{\pm \ast} \eta_1^{\pm} ~\rightarrow ~(q\bar q)~\eta_1^{\pm}, ~(l\nu)~\eta_1^{\pm}.
\label{Eq:n5pp_goff_s}
\eea
At $f=5$ TeV and $C_g \gtrsim 0.01$, the mass of $\eta_{5}^{\pm \pm}$ is $\gtrsim 1.6$ TeV and
the mass difference 
between $\eta_{5}^{\pm \pm}$ and $\eta_{1}^{\pm}$ becomes sufficient to produce an on-shell $W$ boson, 
\bea
\eta_{5}^{\pm \pm}~ &\rightarrow & W^{\pm} \eta_1^{\pm},
\label{Eq:n5pp_gon_s}
\eea
and this decay mode dominates for $m_{\eta_{5}^{\pm \pm}} \gtrsim 1.6$ TeV.
The couplings of the pNGB scalars 
to the gauge bosons are suppressed because these couplings are generated via five dimensional operators.
Hence, the branching ratio for 
\bea
\eta_{5}^{\pm \pm}~ &\rightarrow & W^{\pm} W^{\pm},
\label{Eq:n5pp_gg}
\eea
is suppressed. 
In principle, the decay via off-shell pNGB \emph{i.e.,} $\eta_{5}^{\pm \pm}\rightarrow W^{\pm} \eta_{i}^{\pm \ast}$, $i=1,2$ and $\eta_{5}^{\pm \pm}\rightarrow Z \eta_5^{\pm\pm*}$ 
are also possible, leading to three gauge bosons in the final states. However the decay width of these oﬀ shell scalars to the gauge bosons are suppressed.

The fermiophilic branching ratios for $\eta_{5}^{\pm \pm}$ are depicted in Fig.~\ref{fig:n5pp} (bottom panel). 
The decay of $\eta_{5}^{\pm\pm}$ to the gauge bosons are suppressed as before. 
At $f=1$ TeV, for smaller masses, the three-body decay of the fermiophobic case (Eq.~\ref{Eq:n5pp_goff_s}) is dominant \emph{i.e.,}
\bea
\eta_{5}^{\pm \pm}~ \rightarrow  W^{\pm \ast} \eta_1^{\pm}\rightarrow (q\bar q)\eta_1^{\pm},(l\nu)\eta_1^{\pm}. \nonumber
\eea
With increase in mass, however, the fermiophilic three-body decay to one gauge boson and two SM fermions via an off-shell pNGB\footnote {Here, in the writing, we have used the redefinition $t= t, \bar t$, $b=b, \bar b$, $q=q, \bar q$ and $l=l^+,l^-$, where particle and antiparticle are summed over.} \emph{i.e.,}
\bea
\eta_{5}^{\pm \pm}~ \rightarrow W^{\pm} \eta_{i}^{\pm \ast} \rightarrow W^{\pm} t b,
\label{Eq:n5pp_wtb}
\eea
where $i=1,2$, becomes feasible and is the dominating decay mode 
for most of the parameter space\footnote{The doubly charged pNGB does not have any direct coupling with the SM quarks and therefore, two body decays of $\eta_{5}^{\pm \pm}$ to two SM quarks are absent.}. 
However, with increasing \emph{f}, 
the branching ratios of the pNGB scalars to the fermions get smaller since the fermiophilic coupling is proportional to $m_t/f$.
Hence, at larger masses, when $f=5$ TeV and $C_g \gtrsim 0.01$, 
the two body on-shell decay \emph{viz.,}
\bea
\eta_{5}^{\pm \pm}~ \rightarrow W^{\pm} \eta_{1}^{\pm},
\eea 
comes to dominate.

The three-body decays where the off-shell pNGB scalars decay to two gauge bosons are suppressed since coupling of the pNGB scalars with the SM gauge bosons occurs via dimension-5 terms. 
On the other hand, pNGB scalars coupling with the fermions occurs at tree level and hence, is not suppressed. The three-body decay to $W^{\pm} t b$ can also proceed via $W^{\pm}W^{\pm*}$ 
where the off-shell gauge boson decays to two SM fermions. However, this contribution is negligible compared to the fermiophilic three-body decay via the off-shell pNGB. 

Therefore, in the decays of $\eta_{5}^{\pm \pm}$, the common decay products are the singly charged pNGBs. 
Next, we discuss the decays of these singly charged pNGBs.
\subsection{Branching Ratios of $\eta_2^{\pm}$}
\begin{figure}[h!]
\includegraphics[height=5.2cm,width=5.7cm]{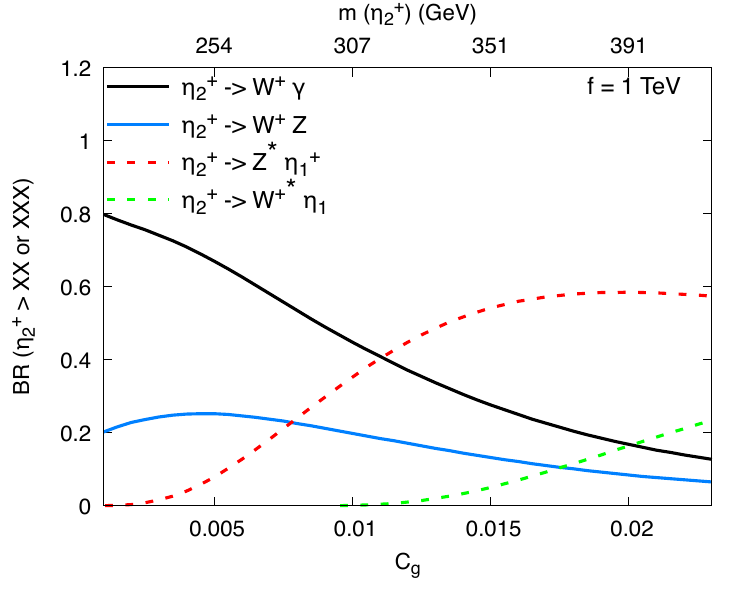}
\includegraphics[height=5.2cm,width=5.7cm]{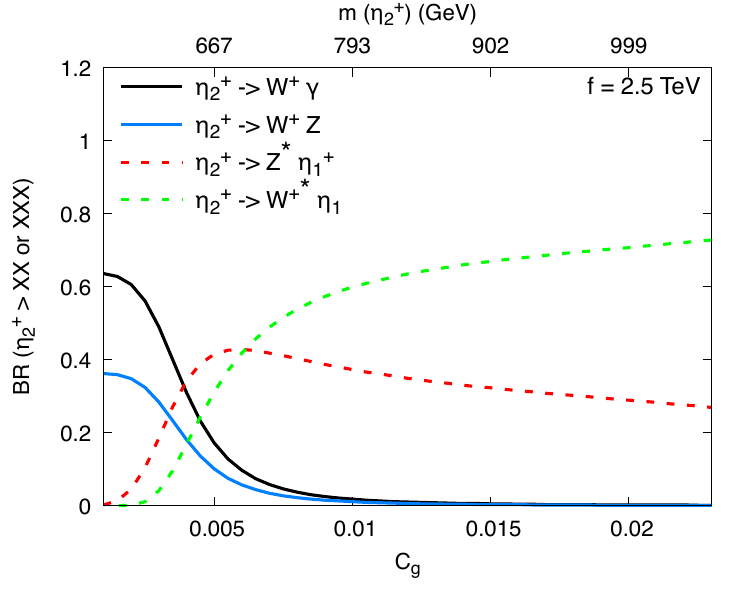}
\includegraphics[height=5.2cm,width=5.7cm]{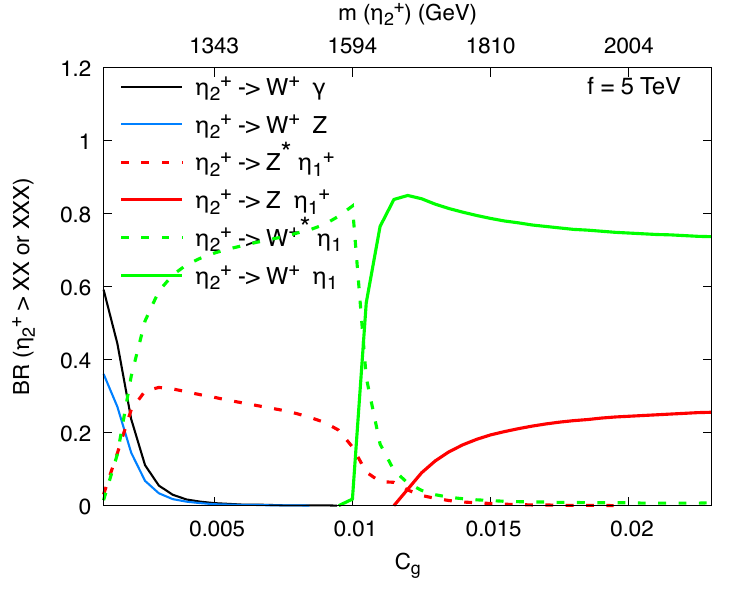}   \\       
\includegraphics[height=5.2cm,width=5.7cm]{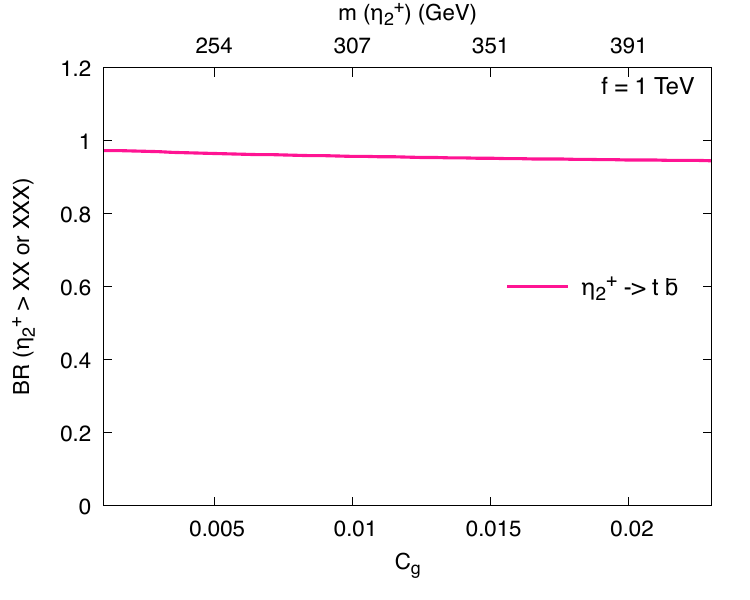}
\includegraphics[height=5.2cm,width=5.7cm]{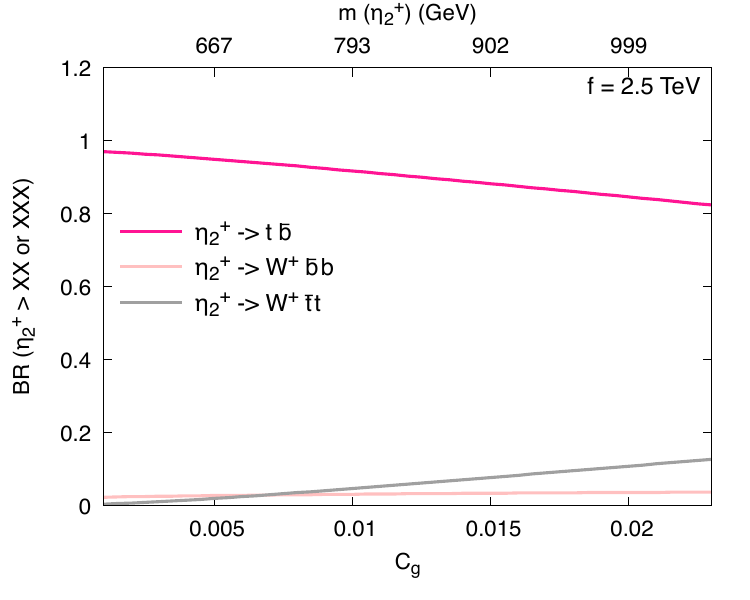}
\includegraphics[height=5.2cm,width=5.7cm]{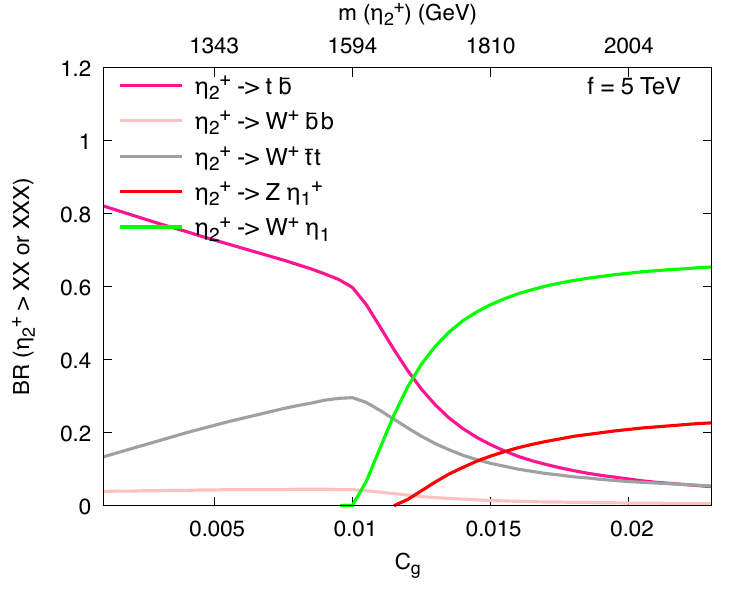} 
    \caption{Branching ratios of $\eta_{2}^{\pm}$ for the \emph{fermiophobic} scenario (top panel) and 
    the \emph{fermiophilic} scenario (bottom panel) at \emph{f} : 1 TeV (left), 2.5 TeV (middle) and 5 TeV (right).}
    \label{fig:n2p}
    \end{figure}

The branching ratios of the singly charged pNGB scalar $\eta_{2}^{\pm}$ are plotted in Fig.~\ref{fig:n2p}, 
for the fermiophobic scenario (top panel) and the fermiophilic scenario (bottom panel). The various parameters are the same 
as before.
First, we discuss the \emph{fermiophobic} case.
The main decay channels of $\eta_2^{\pm}$ up to $f=2.5$ TeV are,
\bea
\eta_2^{\pm}~ &\rightarrow & W^{\pm} Z ,  W^{\pm} \gamma, \n \\
\eta_2^{\pm}~ &\rightarrow & Z^{\ast} \eta_1^{\pm} \rightarrow (ll)\eta_1^{\pm},(q\bar q)\eta_1^{\pm}, \n \\
\eta_2^{\pm}~ &\rightarrow & W^{\pm \ast} \eta_1 \rightarrow (l\nu)\eta_1,(q\bar q)\eta_1. 
\eea
At very small masses, 
in the absence of sufficient mass-splitting with the pNGB scalars, $W^{\pm} \gamma$ emerges as the dominant decay mode. 
As the mass increases, decays to $\eta_1$ and $\eta_1^{\pm}$ \footnote{Considering the mass-hierarchy, decay of $\eta_2^{\pm}$ 
to any other pNGB scalar is not allowed.} become possible accompanied by off-shell gauge bosons 
and these decay modes start to dominate, which is shown in Fig.~\ref{fig:n2p} (top left and middle).
With increase in \emph{f}, the coupling of $\eta_2^{\pm}$ with $\eta_1$ 
is enhanced compared to the coupling with $\eta_1^{\pm}$.
Effect of this in the branching ratio is depicted by the cross over of the red and green dashed curves in Fig.~\ref{fig:n2p} (middle). 
With further increase in \emph{f}, shown by the top right plot in Fig.~\ref{fig:n2p} (\emph{f}$=5$ TeV), 
the same decays proceed with on-shell gauge bosons as mass-splittings among the scalars are increased \emph{i.e.,}
\bea
\eta_2^{\pm}~ &\rightarrow & Z ~\eta_1^{\pm}, \n \\
\eta_2^{\pm}~ &\rightarrow & W^{\pm} \eta_1.
\eea
Further, it is found that decays of $\eta_2^{\pm}$ to $\eta_3^{0}$ and $\eta_5^{\pm\pm}$, accompanied by a gauge boson, are suppressed throughout the parameter space.

If we allow for the fermiophilic couplings, decays of $\eta_{2}^{\pm}$ proceed as shown in Fig.~\ref{fig:n2p} (bottom panel). 
For most of the parameter space, especially at $f=1$ and $2.5$ TeV, the two-body fermiophilic decay,
\bea
\eta_2^{\pm}~&\rightarrow& t b, 
\eea
emerges as the dominant decay mode, subjugating all the \emph{fermiophobic} decay modes. Further, with increase in mass, 
the following three-body decays via off-shell pNGB scalar or SM quark also arise
\bea
\eta_2^{\pm}~&\rightarrow& W^{\pm} \eta_{i}^{\ast} \rightarrow W^{\pm} t t, W^{\pm} b b
\eea
where $\eta_{i}^{\ast}$ represents all the neutral scalars \emph{viz.,} $\eta_1, \eta_2$ and $\eta_3^0$. 
The branching ratio to $W^{\pm} t t$ increase as more phase space becomes available for them however branching ratio to $W^{\pm} b b$ is small 
if we consider only the off shell scalar pNGB in the intermediate state\footnote {It is interesting to note that 
$W^{\pm} b b$ decay also occurs via off-shell quark($t^\ast$) and this contribution is larger than the contribution via off-shell pNGB because $\eta_2^\pm$ coupling with quarks is 
enhanced due to mixing (See Table \ref{table:coup_fermion}). However we did not include the contribution from the off shell SM particles while calculating the 
contributions of the off shell scalars in the same final state. This applies to the rest of the studies as well.}.
Hence, these branching ratios are much smaller compared to the two body final state $tb$.
The branching ratios for $\eta_2^{\pm}~ \rightarrow  W^{\pm\ast} \eta_{i}$, $\eta_2^{\pm}\rightarrow W^{\pm\ast} \eta^{\mp\mp}$ and $\eta_2^{\pm}~ \rightarrow  Z \eta_{1}^{\pm\ast}
\rightarrow  Z tb$ are found to be very small. The Branching ratio to the off-shell $\eta_5^{\pm\pm}$ is also very small as $\eta_5^{\pm\pm}$ does not couple directly to the fermions.

The branching ratios for $\eta_2^\pm$ decays to the fermions fall with increase in \emph{f} as the involved coupling is 
inversely proportional to \emph{f} (see Fig.~\ref{fig:n2p}, bottom right). 
At $f=5$ TeV and $C_g \gtrsim 0.01$, the dominating mode for $\eta_2^\pm$ decay is to 
on-shell gauge bosons and pNGB scalars \emph{i.e.,}
\bea
\eta_2^{\pm}~ \rightarrow  W^{\pm} \eta_{1},\\
\eta_2^{\pm}~ \rightarrow Z \eta_1^{\pm}.
\eea
\subsection{Branching Ratios of $\eta_1^{\pm}$}
\begin{figure}[h!]
\includegraphics[height=5.2cm,width=5.7cm]{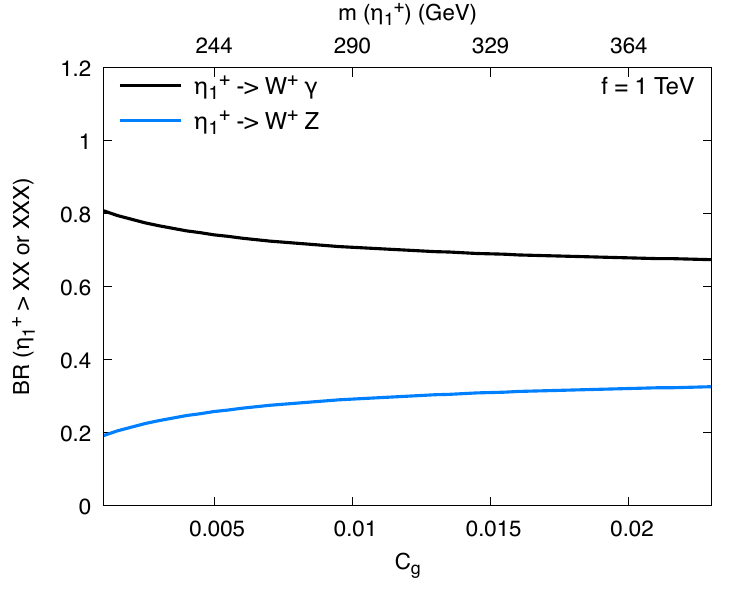}
\includegraphics[height=5.2cm,width=5.7cm]{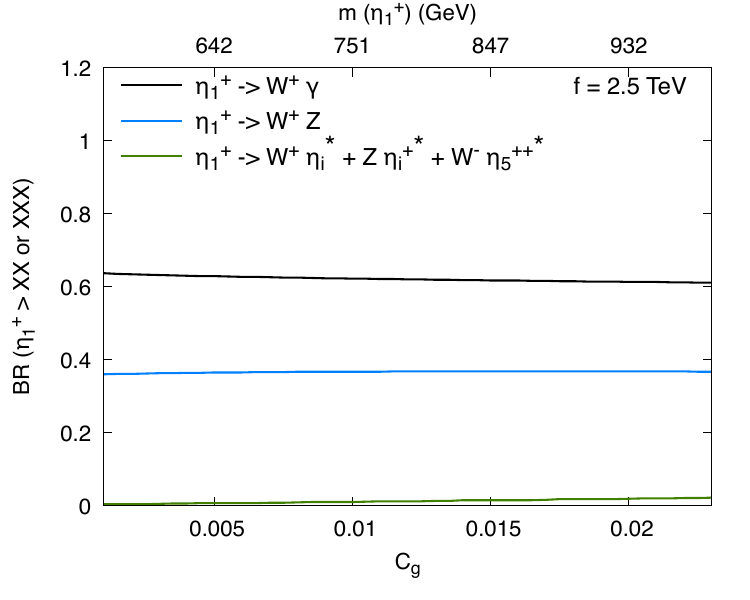}
\includegraphics[height=5.2cm,width=5.7cm]{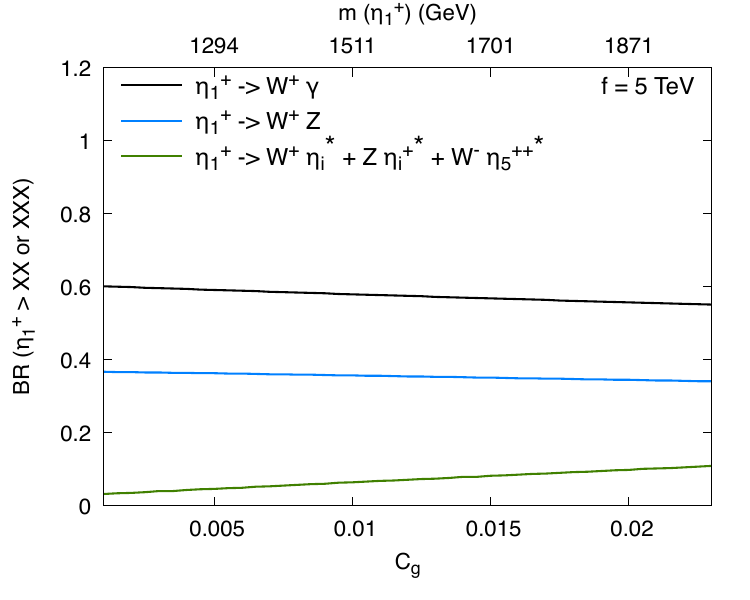}   \\       
\includegraphics[height=5.2cm,width=5.7cm]{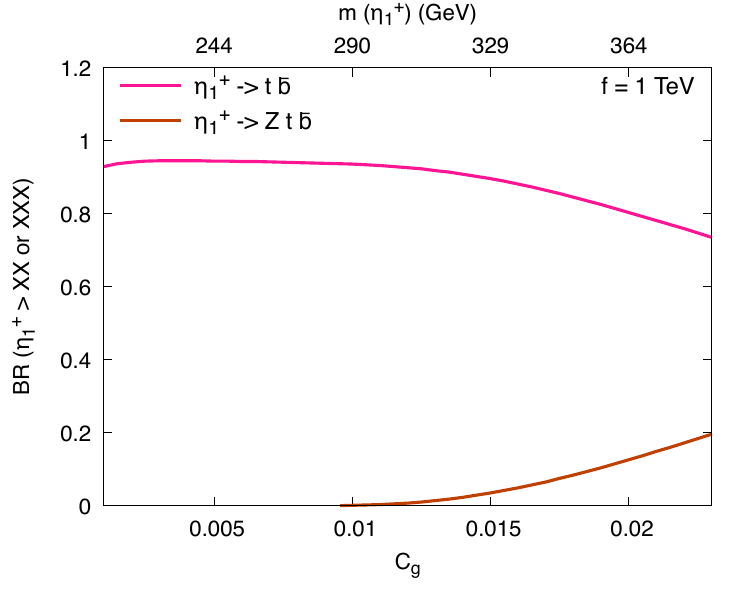}
\includegraphics[height=5.2cm,width=5.7cm]{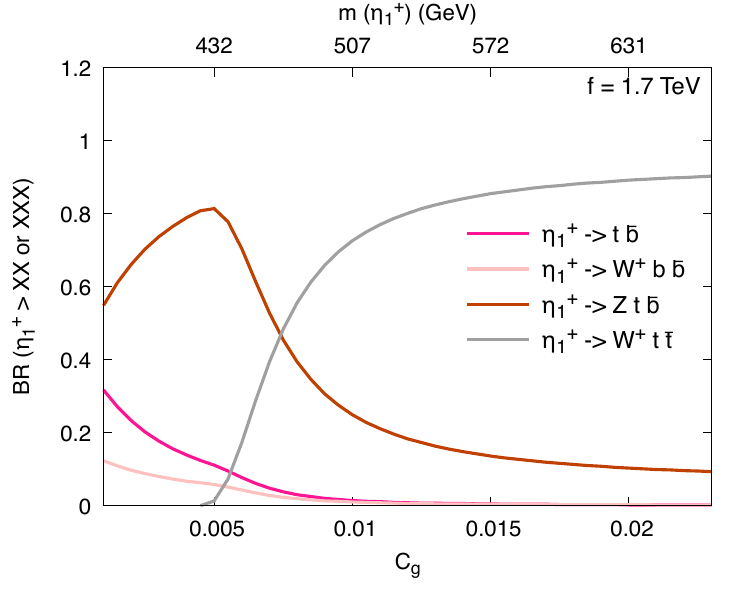}
\includegraphics[height=5.2cm,width=5.7cm]{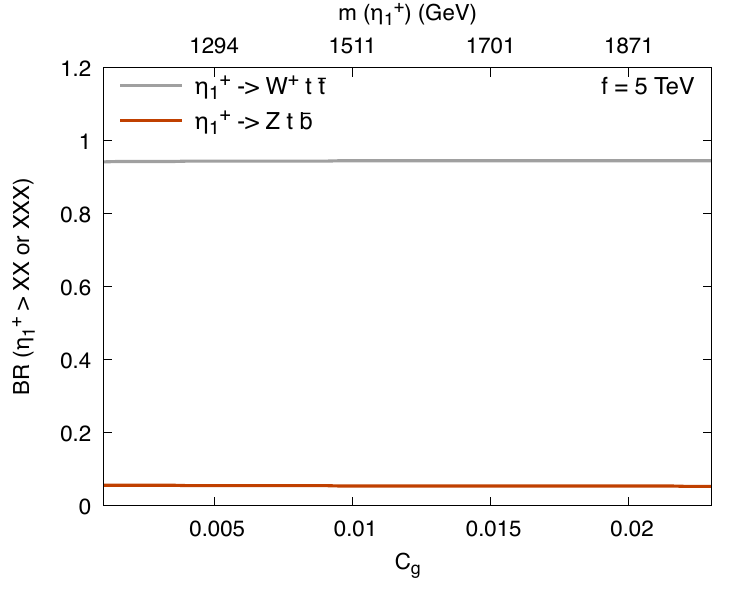}
   \caption{Branching ratios of $\eta_{1}^{\pm}$ for the \emph{fermiophobic} scenario (top panel) and 
   the \emph{fermiophilic} scenario (bottom panel) at \emph{f} : 1 TeV (left), 2.5 TeV (1.7 TeV for \emph{fermiophilic}) (middle) and 5 TeV (right).}
    \label{fig:n1p}
    \end{figure}

The branching ratios of $\eta_{1}^{\pm}$ are very interesting as they differ from that of the other singly charged scalar, $\eta_{2}^{\pm}$.
The graphs are plotted in Fig.~\ref{fig:n1p}, for the fermiophobic scenario (top panel) and the fermiophilic scenario (bottom panel).

First, we discuss the fermiophobic scenario. From Table \ref{table:masses}, we can observe that at 
lower values of $C_g$, $\eta_1^{\pm}$ has the lowest mass in the spectrum and therefore, any on-shell decays 
involving other pNGB and SM gauge bosons are prohibited. At larger $C_g$, $\eta_1^{\pm}$ is the second lightest pNGB scalar, 
next to the lightest, $\eta$, but $\eta$ does not couple to $\eta_1^{\pm}$ or any other pNGB scalars. 
As a result, $\eta_1^{\pm}$ cannot decay to an on-shell pNGB scalar, at all, which is unlike the case of $\eta_2^{\pm}$.
Instead, it prefers to decay to gauge bosons. The decay modes that are dominant at smaller masses ($f=1$ and $2.5$ TeV) are,
\bea
\eta_1^{\pm}~~&\rightarrow & W^{\pm} Z ,  W^{\pm} \gamma,
\eea
as shown in Fig.~\ref{fig:n1p} (top left and middle). 
At large masses ($f=5$ TeV, Fig.~\ref{fig:n1p} (top right)), 3-body decays via off-shell pNGB scalars become significant \emph{i.e.,}
\bea
\label{Eq:3gauge_scalar}
\eta_1^{\pm}~~&\rightarrow & W^{\pm} \eta_{i}^{\ast}, \\
\eta_1^{\pm}~~&\rightarrow & W^{\mp} \eta_{5}^{\pm\pm \ast}, \\
\eta_1^{\pm}~~&\rightarrow & Z \eta_{2}^{\pm\ast},
\eea
where $\eta_i=\eta_1,~\eta_2$ or $\eta_3^0$. These off-shell scalars finally decay to the gauge bosons leading to 
final states with three gauge bosons.
However branching ratios in these channels are suppressed, as shown in Fig.~\ref{fig:n1p} (top right).
Hence, we observe that in the fermiophobic case, the decay patterns of  $\eta_1^{\pm}$ and $\eta_2^{\pm}$ are very different. 

We discuss the fermiophilic decays of $\eta_1^{\pm}$, now.
The dominating decay modes at $f =$ 1 TeV is,
\bea
\eta_1^{\pm}~&\rightarrow&  t b
\eea
which is shown in Fig.~\ref{fig:n1p} (bottom left). 
In Fig.~\ref{fig:n1p} (bottom middle), we have plotted the branching ratios for $f=1.7$ TeV. 
Unlike other pNGB scalar plots, we choose $f=1.7$ TeV here for better depiction of the variation in branching ratios for intermediate masses. 
We find that in a very small region of mass range, 
\bea
\eta_1^{\pm}~&\rightarrow& Z \eta_{2}^{\pm\ast} \rightarrow Z t b
\eea
dominates. This is because the $\eta_1^{\pm} tb$ coupling is suppressed compared to $\eta_2^{\pm} tb$ (see Table \ref{table:coup_fermion}). 
Hence, this is unlike the case of $\eta_2^{\pm}$ where due to its enhanced coupling with SM quarks, 
the decay to $tb$ dominates the decay mode $Ztb$. The three-body branching ratio to the final state $W^{\pm} b b$ is found to be very small throughout the parameter space.
The three-body decays via off-shell $\eta_5^{\pm\pm}$ are absent since $\eta_5^{\pm\pm}$ does not couple directly to the fermions.

With further increase in the mass of $\eta_1^{\pm}$, the process
\bea
\eta_1^{\pm}~ \rightarrow  W^{\pm} \eta_{i}^{\ast} \rightarrow W^{\pm} t\bar{t}
\eea
takes over and dominates its branching ratio, especially at $f=5$ TeV (Fig.~\ref{fig:n1p}, bottom right).
As discussed before, the mass of $\eta_1^{\pm}$ is always less than the mass of $\eta_2^{\pm}$ and $\eta_{i}$. 
Hence, we do not observe any on-shell decay of $\eta_1^{\pm}$ to other pNGB scalar, especially at large values of \emph{f}. 
The absence of these decays in $\eta_1^{\pm}$ can be useful to distinguish the signature of $\eta_1^{\pm}$ from $\eta_2^{\pm}$ and vice-versa 
at the collider experiments.
\subsection{Branching Ratios of the Singlet Neutral scalar, $\eta$}
\begin{figure}[h!]
\includegraphics[height=5.2cm,width=5.7cm]{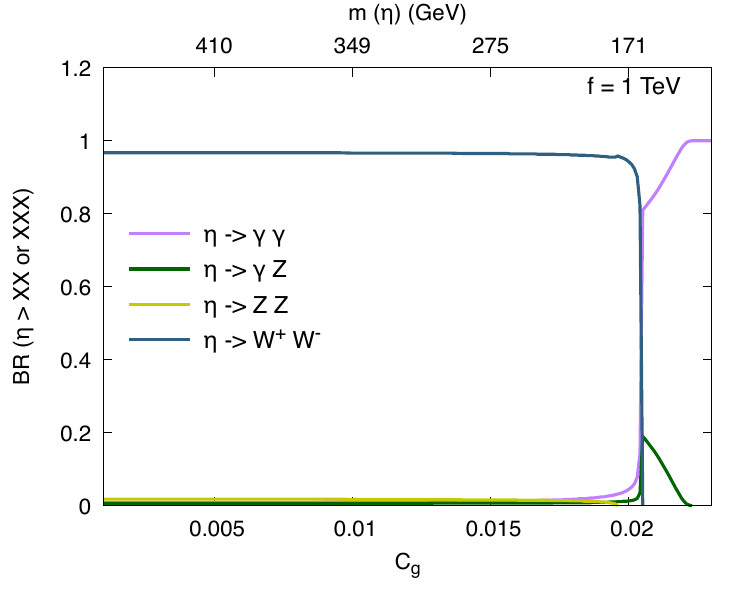}
\includegraphics[height=5.2cm,width=5.7cm]{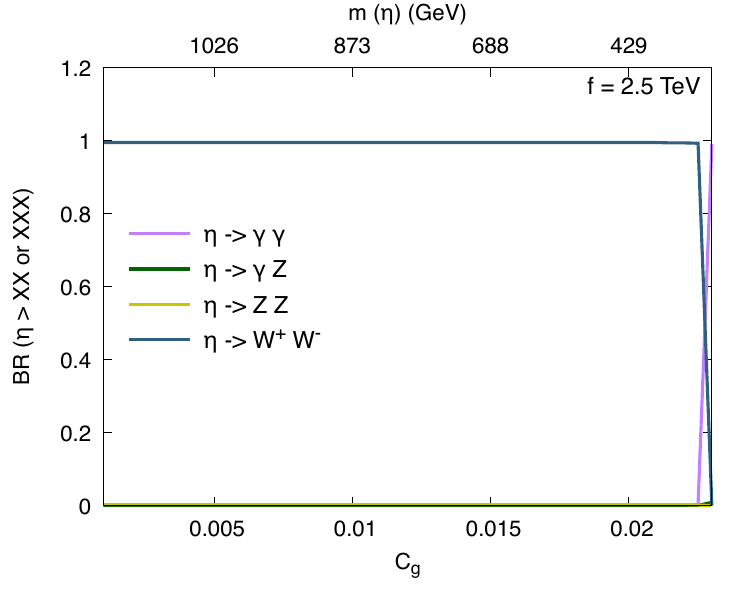}
\includegraphics[height=5.2cm,width=5.7cm]{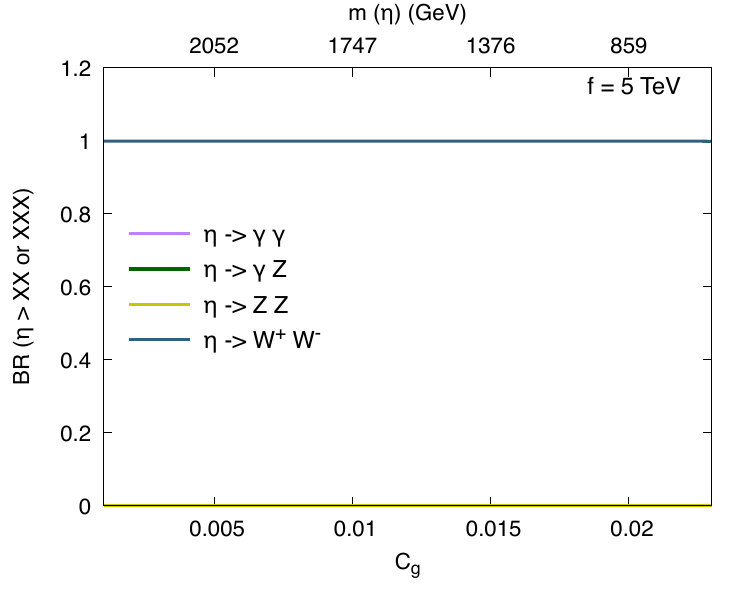} \\
\includegraphics[height=5.2cm,width=5.7cm]{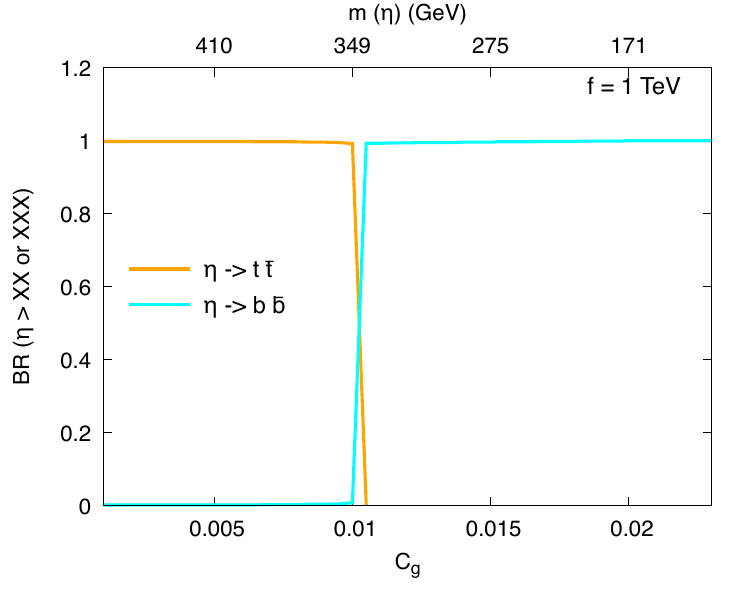}
\includegraphics[height=5.2cm,width=5.7cm]{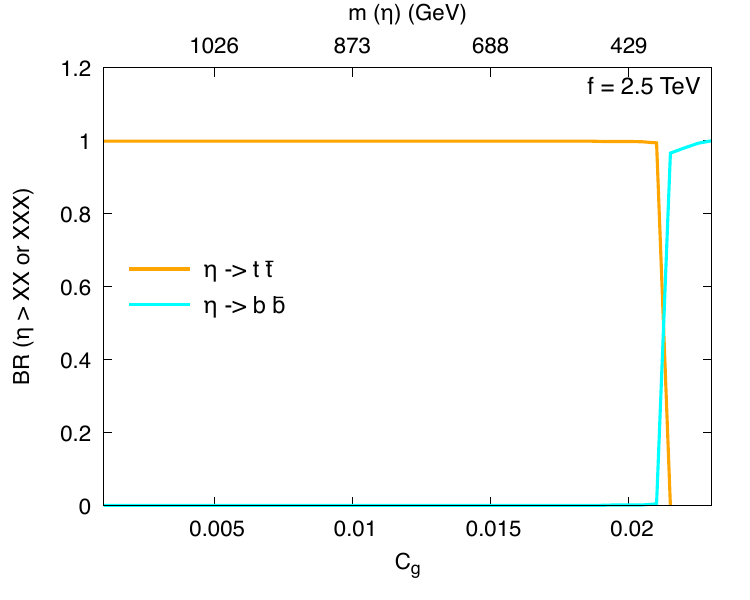}
\includegraphics[height=5.2cm,width=5.7cm]{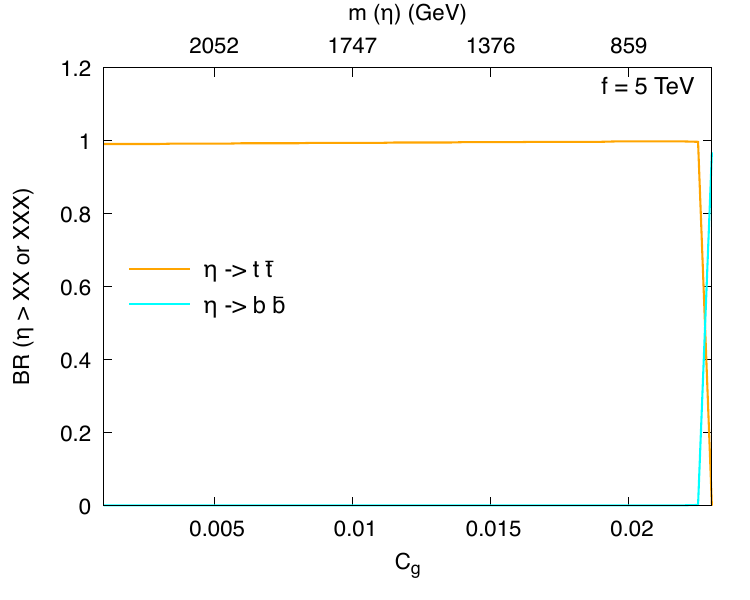}
    \caption{Branching ratios of the singlet scalar $\eta$ for the \emph{fermiophobic} scenario (top panel) and 
   the \emph{fermiophilic} scenario (bottom panel) at \emph{f} : 1 TeV (left), 2.5 TeV (middle) and 5 TeV (right).}
    \label{fig:n00}
    \end{figure}

The branching ratios of the neutral pNGB scalars are also very interesting which we discuss next.
The branching ratios for the singlet scalar, $\eta$, are plotted in Fig.~\ref{fig:n00}, for the fermiophobic
scenario (top panel) and the fermiophilic scenario (bottom panel).
Earlier, we discussed that the singlet scalar $\eta$ couples to the gauge bosons but it does not have couplings to any other pNGB scalars. Hence, the decay modes of this neutral scalar, in the fermiophobic scenario, are:
\bea
\eta~~\rightarrow  ~~\gamma Z , \gamma \gamma, ZZ, W^{+} W^{-},\n
\eea
as shown in Fig.~\ref{fig:n00} (top panel). The dominating decay mode is $W^{+} W^{-}$, 
for most of the parameter space. However, towards the upper limit of $C_g$ \emph{i.e.,} when the mass of $\eta$ is small, 
$\gamma \gamma$ becomes the dominating decay mode. In the fermiophilic scenario \emph{i.e.,} 
Fig.~\ref{fig:n00} (bottom panel), the major decay modes are
\bea
\eta~~&\rightarrow &  ~~bb,  tt.
\eea
The preferential decay to $tt$ at small values of $C_g$ and large values of \emph{f} may be easily understood from the dependence 
of $\eta$ mass on these parameters, as depicted on the top axis, parallel x-axis of the plots. As the mass of $\eta$ decreases 
with the increase of $C_g$, at smaller values of $C_g$, decay to $tt$ dominates whereas, at larger values of $C_g$, 
decay to $bb$ dominates when $f=1$ TeV. However, as the compositeness scale \emph{f} increases, the mass also increases 
for a fixed $C_g$. Hence, at large \emph{f}, decay to $tt$ starts to dominate again.
As for the cases of other pNGB scalars, here as well, decays to the gauge bosons are suppressed in the fermiophilic scenario.
\subsection{Branching Ratios of $\eta_2$ }
\begin{figure}[h!]
\includegraphics[height=5.2cm,width=5.7cm]{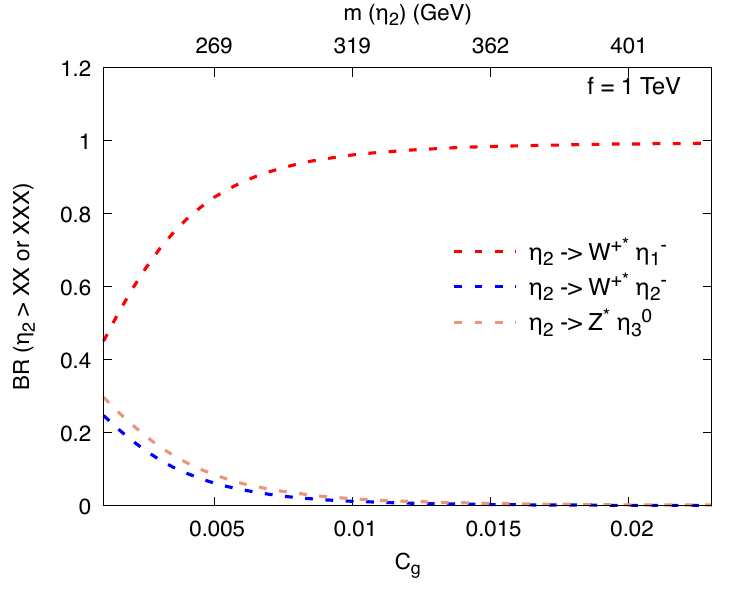}
\includegraphics[height=5.2cm,width=5.7cm]{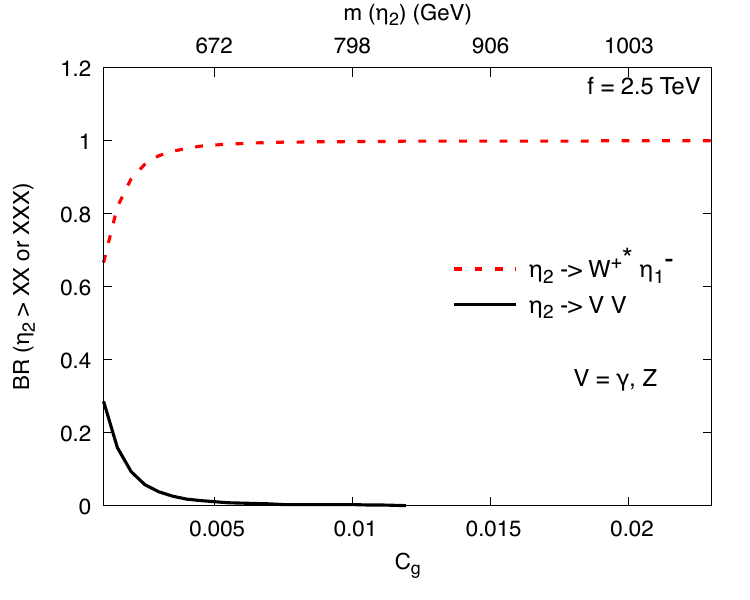}
\includegraphics[height=5.2cm,width=5.7cm]{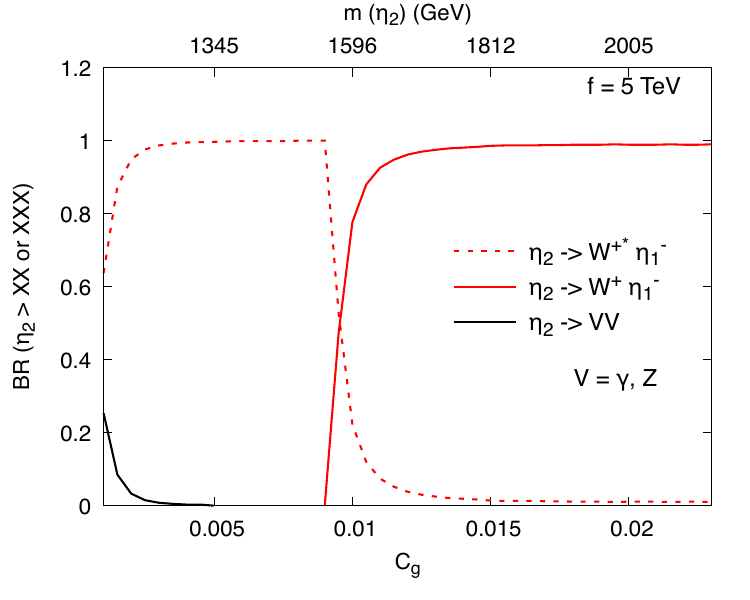} \\
\includegraphics[height=5.2cm,width=5.7cm]{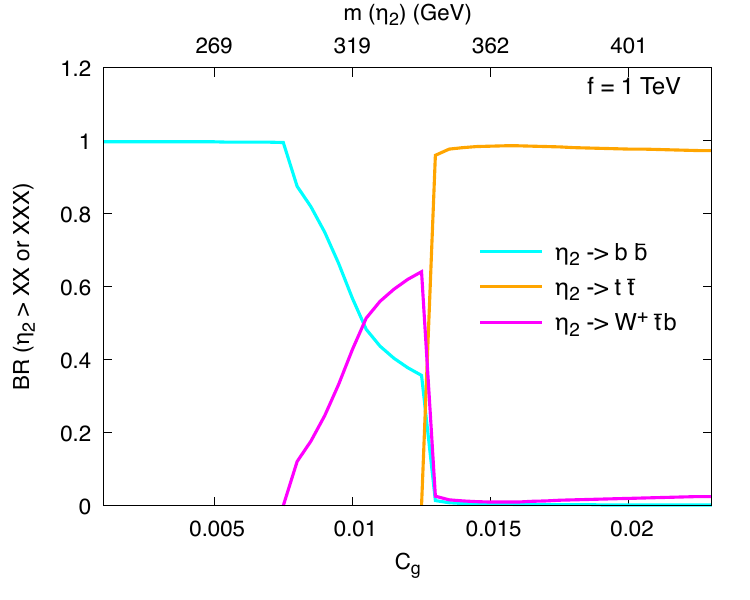}
\includegraphics[height=5.2cm,width=5.7cm]{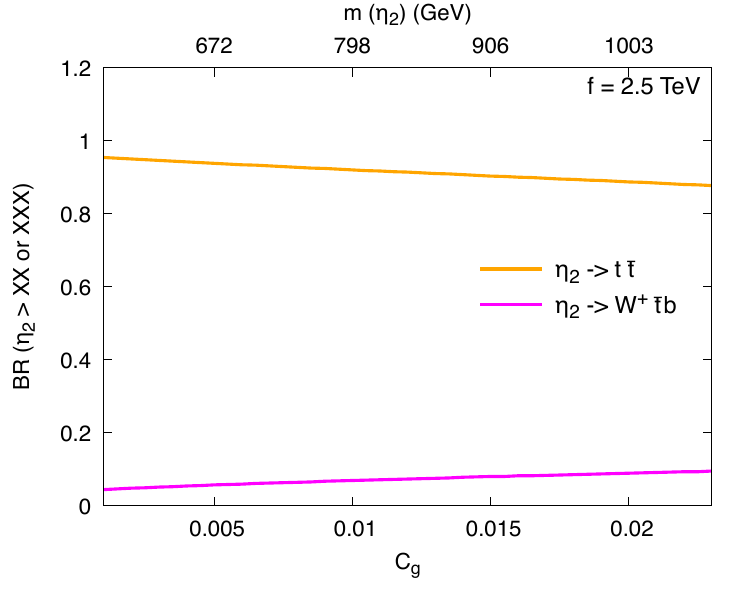}
\includegraphics[height=5.2cm,width=5.7cm]{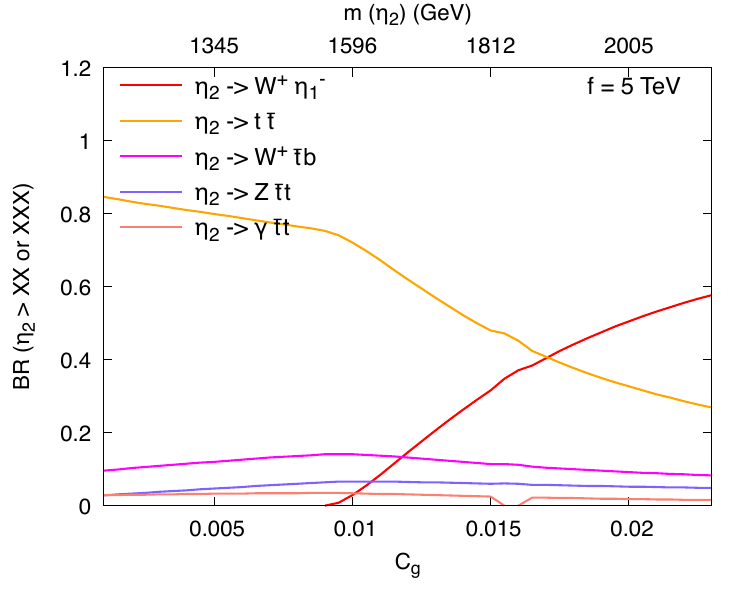}

    \caption{Branching ratios of the singlet scalar $\eta_2$ for the \emph{fermiophobic} scenario (top panel) and 
   \emph{fermiophilic} scenario (bottom panel) at \emph{f} : 1 TeV (left), 2.5 TeV (middle) and 5 TeV (right).}
    \label{fig:n20}
    \end{figure}

The decays of the neutral scalar, $\eta_2$, are shown in Fig.~\ref{fig:n20}, for the \emph{fermiophobic} 
scenario (top panel) and the \emph{fermiophilic} scenario (bottom panel). 
We, first, discuss the fermiophobic scenario (Fig.~\ref{fig:n20}, top panel). Considering the mass-hierarchy 
among the pNGB scalars, the neutral scalar $\eta_2$ decays to lighter pNGB scalars. At small masses, only three-body decays via off-shell $W/Z$ are possible \emph{i.e.,}
\bea
\eta_2 &\rightarrow&  W^{\mp \ast} \eta_{i}^{\pm} \rightarrow (l\nu) \eta_{i}^{\pm}, (qq) \eta_{i}^{\pm}, \\
\eta_2 &\rightarrow&  Z^{\ast} \eta_{3}^{0} \rightarrow (ll) (qq)  \eta_{3}^{0}. 
\eea
These are the dominating modes at $f=$ 1 and 2.5 TeV, with the branching ratio for $\eta_2 \rightarrow W^{\pm \ast} \eta_{1}^{\pm}$ 
being larger due to large mass-splitting between $\eta_2$ and $\eta_1^{\pm}$. The branching ratio to the gauge bosons \emph{viz.,}
\bea
\eta_2~~&\rightarrow & \gamma Z , \gamma \gamma, ZZ, W^{\pm} W^{\mp} 
\eea
are suppressed. 
At $f = 5$ TeV, due to the increased phase space, the on-shell decay $\eta_2 \rightarrow  W^{\pm} \eta_{1}^{\mp}$ 
dominates when $C_g \gtrsim 0.01$. 

In the fermiophilic scenario (Fig.~\ref{fig:n20}, bottom panel), the following decays to SM fermions are dominant,
\bea
\eta_2 &\rightarrow& bb, \\
\eta_2 &\rightarrow& tt, \\
\eta_2 &\rightarrow& W^{\mp} \eta_{i}^{\pm\ast}\rightarrow W^{\mp} tb,
\eea
where, $\eta_2\rightarrow bb$ is dominant at very small masses \emph{viz.,} for $f=1$ TeV and $C_g \lesssim 0.01$. 
With increase in mass ($f=1$ and $2.5$ TeV), the branching ratio for $\eta_2 \rightarrow tt$ become significant (see Fig.~\ref{fig:n20}, bottom left and middle).
Branching ratio of $\eta_2 \rightarrow W^{\pm} tb$ is significant in a small mass window at $f=1$ TeV but is suppressed as soon as decay to $tt$ becomes feasible. The branching ratio of $\eta_2 \rightarrow Ztt$ is negligible.

At large \emph{f} (5 TeV), however, the coupling of pNGB scalars with the SM fermions decreases. 
Hence, in the region, $f=5$ TeV and $C_g \gtrsim 0.01$, the on-shell decay mode $\eta_2 \rightarrow  W^{\pm} \eta_{1}^{\pm}$ becomes the dominant decay channel (see
Fig.~\ref{fig:n20}, bottom right).
\subsection{Branching Ratios of $\eta_1$ }
\begin{figure}[h!]
\includegraphics[height=5.2cm,width=5.7cm]{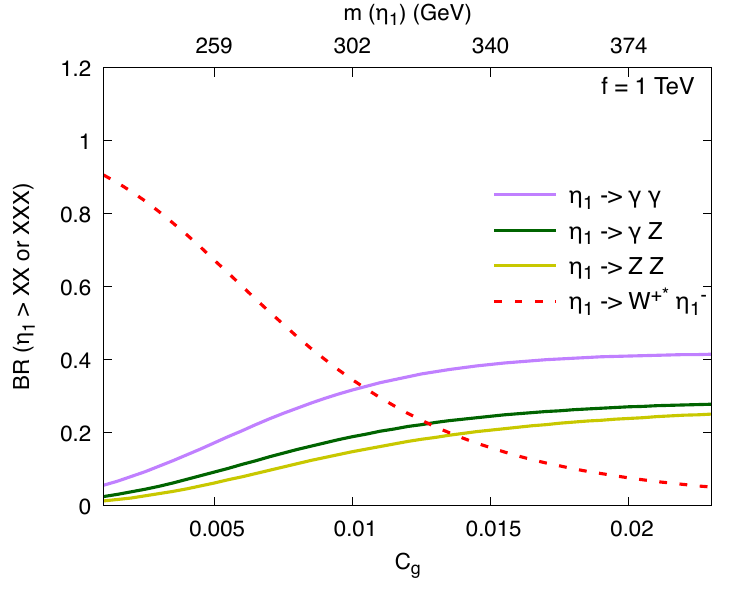}
\includegraphics[height=5.2cm,width=5.7cm]{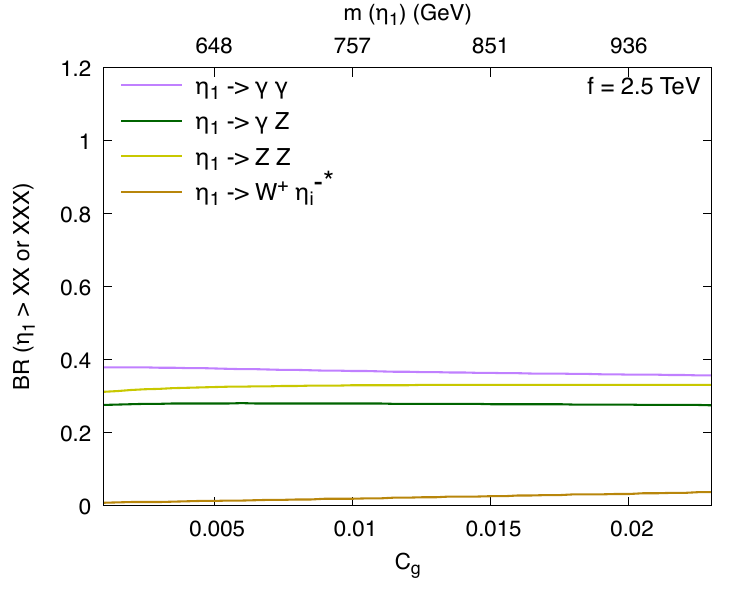}
\includegraphics[height=5.2cm,width=5.7cm]{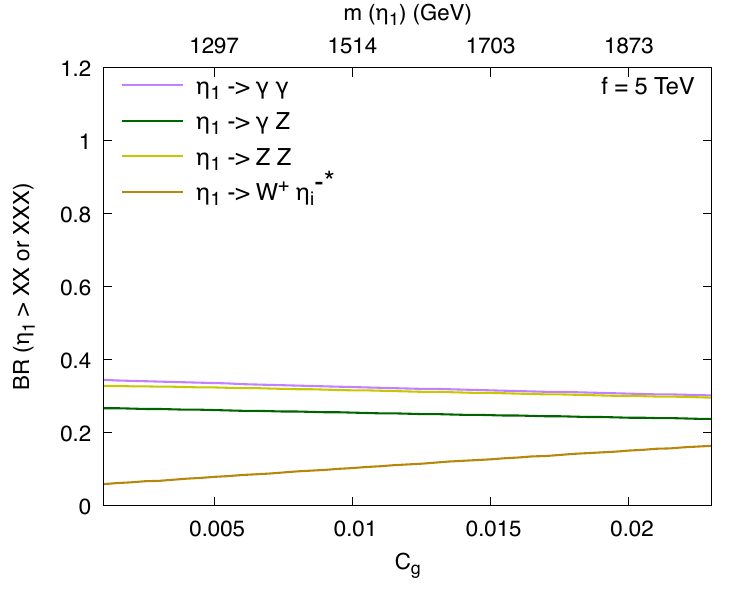} \\
\includegraphics[height=5.2cm,width=5.7cm]{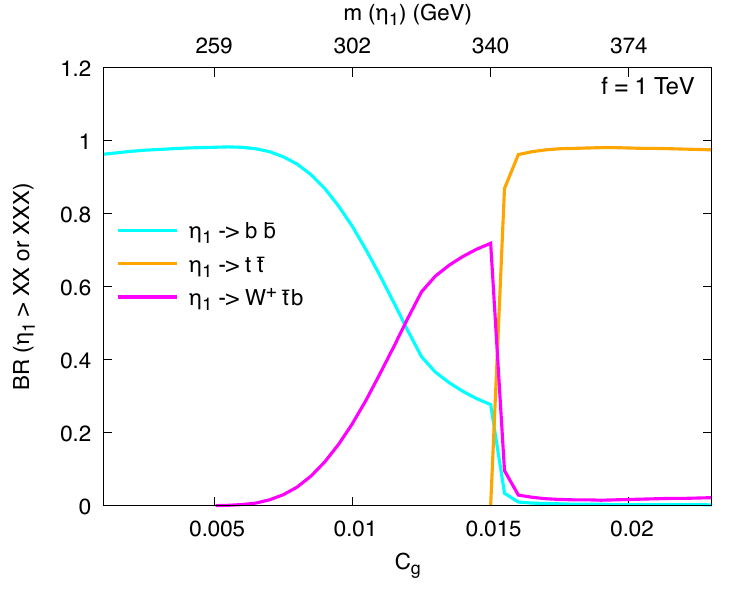}
\includegraphics[height=5.2cm,width=5.7cm]{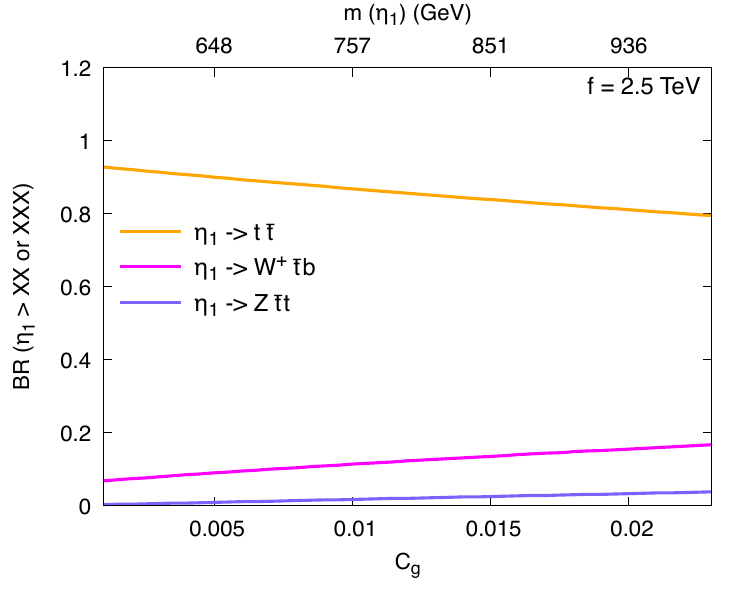}
\includegraphics[height=5.2cm,width=5.7cm]{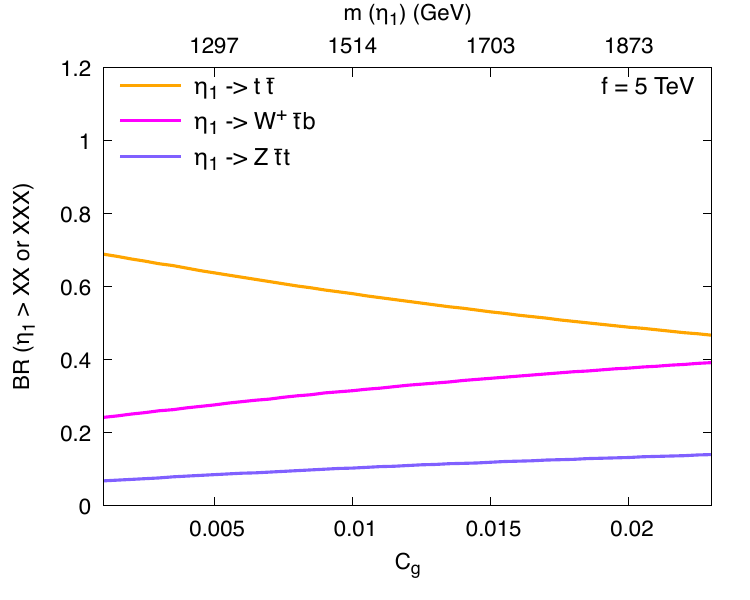}

    \caption{Branching ratios of the singlet scalar $\eta_1$ for the \emph{fermiophobic} scenario (top panel) and 
   \emph{fermiophilic} scenario (bottom panel) at \emph{f} : 1 TeV (left), 2.5 TeV (middle) and 5 TeV (right).}
    \label{fig:n10}
    \end{figure}
Analogous to the case of charged scalars, the neutral scalar, $\eta_1$, exhibits quite a different behaviour than $\eta_2$.
We depict the branching ratios for the fermiophobic scenario in Fig.~\ref{fig:n10} (top panel). 
Considering the mass-hierarchy and mass-splittings among the pNGBs, the only decay of $\eta_1$ to another pNGB scalar is to $\eta_1^{\pm}$, accompanied by the off-shell $W$ boson \emph{i.e.,}  
\bea
\eta_1 \rightarrow  W^{\pm \ast} \eta_{1}^{\pm} \rightarrow (l\nu) \eta_{1}^{\pm}, (qq) \eta_{1}^{\pm}.
\eea
This decay mode dominates at very small masses only because the mass-splitting between $\eta_1$ and $\eta_1^{\pm}$ further diminishes with increase in \emph{f}. 
As the mass increases, the dominant contribution comes from the 
decays to the gauge bosons, mainly from $\gamma \gamma$ (Fig.~\ref{fig:n10}, top left and middle).
In addition to the two-body decays, three-body decays via off-shell singly charged scalars are also possible \emph{viz.,}
\bea
\eta_1 &\rightarrow& W^{\pm} \eta_i^{\mp *}
\eea
where $i=1,2$, as shown in Fig.~\ref{fig:n10} (top middle and right).

In the fermiophilic scenario, the branching ratios of $\eta_{1}$ are almost similar to $\eta_{2}$ with the exception that at large \emph{f}, the on-shell decay 
to $\eta_{1}^{\pm}$ is not possible due to very small mass difference among the pNGB scalars.
\subsection{Branching Ratios of $\eta_3^0$ }
\begin{figure}[h!]
\includegraphics[height=5.2cm,width=5.7cm]{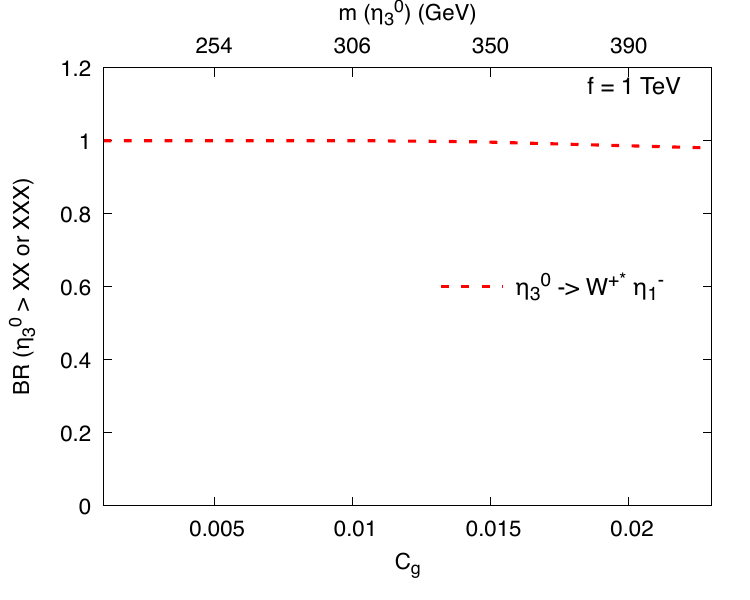}
\includegraphics[height=5.2cm,width=5.7cm]{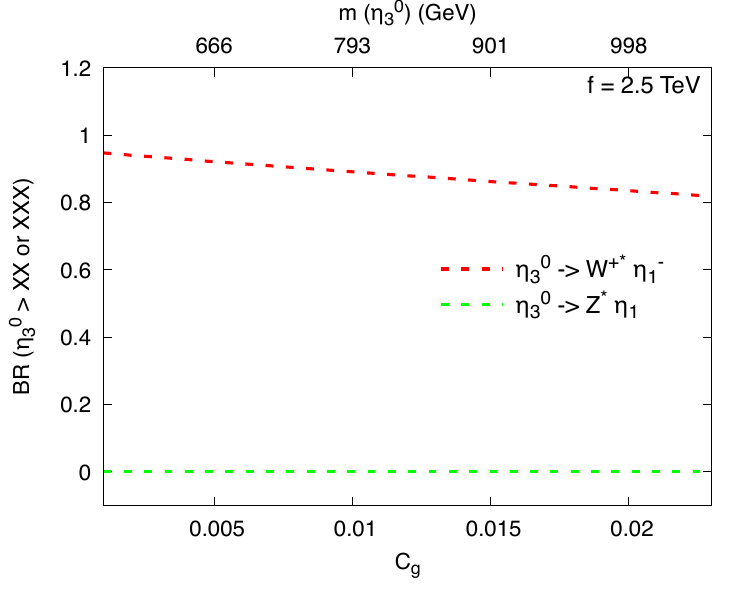}
\includegraphics[height=5.2cm,width=5.7cm]{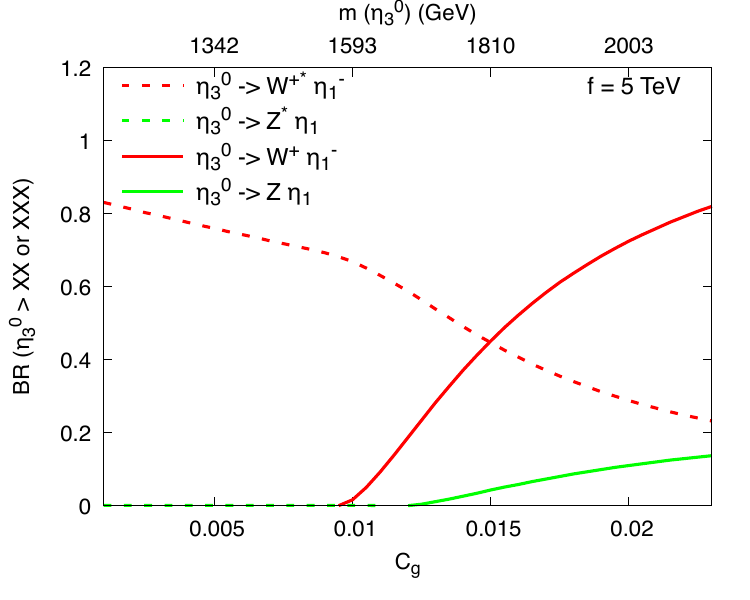} \\
\includegraphics[height=5.2cm,width=5.7cm]{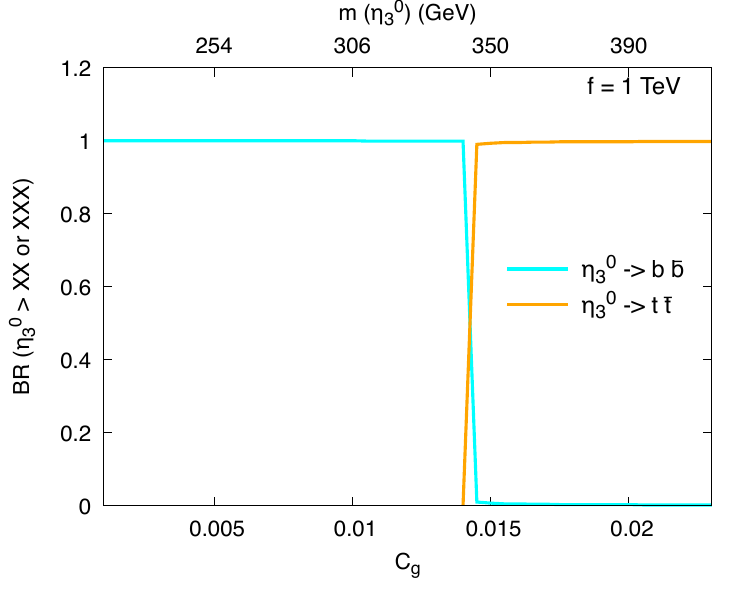}
\includegraphics[height=5.2cm,width=5.7cm]{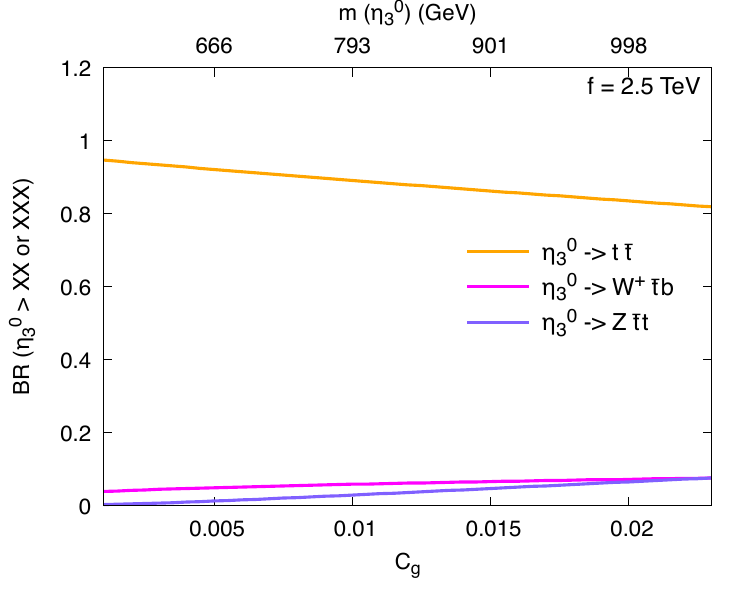}
\includegraphics[height=5.2cm,width=5.7cm]{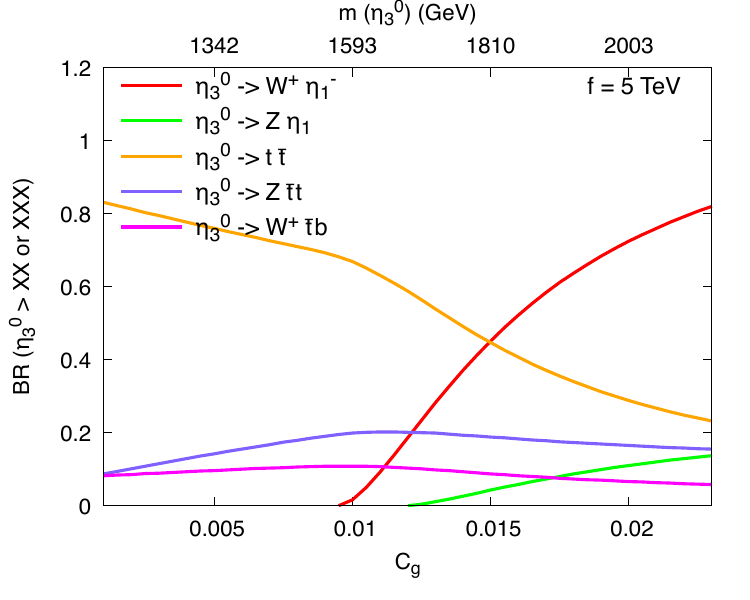}
    \caption{Branching ratios of the singlet scalar $\eta_3^0$ for the \emph{fermiophobic} scenario (top panel) and 
   the \emph{fermiophilic} scenario (bottom panel) at \emph{f} : 1 TeV (left), 2.5 TeV (middle) and 5 TeV (right).}
    \label{fig:n30}
\end{figure}

We, now, discuss the branching ratios of the CP-even scalar, $\eta_3^0$. We start with the fermiophobic scenario (Fig.~\ref{fig:n30}, top panel). 
The neutral pNGB, $\eta_3^0$, does not decay to two gauge bosons. However, it couples to other pNGBs and one gauge boson. The mass of $\eta_3^0$ is close to that of $\eta_2, \eta_2^{\pm}$ and $\eta_5^{\pm\pm}$ and therefore, it can only decay to $\eta_1$ and $\eta_1^{\pm}$. At small masses ($f=1$ and $2.5$ TeV), however, the 
dominating decays are via off-shell gauge bosons resulting into the three-body decays, 
\bea
\eta_3^0 &\rightarrow&  W^{\pm \ast} \eta_{1}^{\mp} \rightarrow (l\nu) \eta_{1}^{\mp}, (qq) \eta_{1}^{\mp}, \\
\eta_3^0 &\rightarrow&  Z^{\ast} \eta_{1} \rightarrow (ll) \eta_{1}, (qq) \eta_{1}.
\eea
For most of the parameter space, the decay mode $\eta_3^0 \rightarrow  W^{\pm*} \eta_{1}^{\mp} + \emph{H.c.}$ dominates (Fig.~\ref{fig:n30}, top left and middle). 
At large \emph{f} ($f=5$ TeV), however, phase space becomes available for the fully on-shell decays,
\bea
\eta_3^0 &\rightarrow&  W^{\pm} \eta_{1}^{\mp},\\
\eta_3^0 &\rightarrow&  Z ~\eta_{1}.
\eea
The decay $\eta_3^0 \rightarrow  W^{\pm} \eta_{1}^{\mp}$ comes to dominate in the region $C_g \gtrsim 0.01$ (Fig.~\ref{fig:n30}, top right).
The branching ratios in the \emph{fermiophilic} case are shown in Fig.~\ref{fig:n30} (bottom panel). The decay patterns are similar to 
that of $\eta_2$ where decays to SM fermions dominate at small \emph{f} while at large \emph{f}, due to reduced coupling to fermions, on-shell decays to lighter pNGB and gauge boson 
dominate. For $\eta_1$, decay to another on-shell pNGB is not possible due to the mass hierarchy. 
\section{Collider Signatures of the Exotic scalars}
The collider signatures of pNGB scalars depend crucially upon their decay modes. For both charged and neutral scalars, their decays vary markedly depending upon whether their couplings with fermions are present or absent vis-\'{a}-vis the fermiophilic and fermiophobic scenarios.
Working in the mass-basis, we derived the masses and mixing 
of the pNGBs, as a function of the model parameters, \emph{f} and $C_g$ and mixing angles, $\kappa_+$ and $\kappa_0$.
Consequently, the branching ratios of the scalars, which are set by the mass spectrum and mixing structure, and are shown in 
Fig.~\ref{fig:n5pp} to Fig.~\ref{fig:n30} as functions of \emph{f} and $C_g$, necessarily inherit a dependence on the mixing angles $\kappa_+$ and $\kappa_0$.
The decay modes of the pNGB scalars along with possible collider signatures were also discussed in \cite{Cacciapaglia:2022bax}, with the mass of the decaying scalar set to $600$ GeV while other scalars were taken to be lighter or heavier.
There, particular emphasis was placed on the three-body decay channels mediated by off-shell gauge bosons, which occur in the regime where the scalar mass splittings lie below the $W$ and $Z$ boson masses.
In this work, however, we explore a wider region of the ($C_g$, $f$) parameter space such that the masses extend up to $\sim 2$ TeV, with correspondingly larger mass splittings.
As a result, we not only observe the above three-body decays but also two-body decays where one of the decay products is a lighter on-shell pNGB scalar. For instance, in Fig.~\ref{fig:n20} (top panel), at $f=1$ and $2.5$ TeV, $\eta_2$ preferentially decays to $\eta_1^{-} W^{+*}$, eventually leading to a three-body decay. At $f=5$ TeV, however, there is sufficient mass-splitting with $\eta_1^{-}$ that the two-body decay, $\eta_2 \rightarrow \eta_1^{-} W^{+}$, can occur and eventually dominate.

We focus on this unexplored parameter space where the pNGB scalars decay to an on-shell gauge boson and another pNGB scalar, in fermiophobic as well as fermiophilic scenarios. The daughter pNGBs continue to decay until the lightest pNGB finally decays into two gauge bosons, resulting in a cascade decay.
In Tables \ref{table:decay1} and \ref{table:decay2}, we express the branching ratios for a few of these decays, for the fermiophobic and fermiophilic cases, respectively. 
These decay modes lead to final states rich in gauge bosons, with large branching ratios. Hence, depending on the subsequent decays of the $W/Z$ bosons, multi-lepton and/or multi-jet final states can be observed at the collider experiments.
\begin{table}[h!]
\centering
\begin{tabular}{|c|c|}
\hline
{\bf Cascade Decay} & {\bf Branching Ratio} \\
\hline
$\eta_5^{\pm \pm}\rightarrow W^{\pm} \eta_1^{\pm} \rightarrow  W^{\pm}  W^{\pm} \gamma$ & $0.33-0.55$\\
$\eta_5^{\pm \pm}\rightarrow  W^{\pm} \eta_1^{\pm} \rightarrow  W^{\pm}  W^{\pm} Z$ & $0.20-0.34$\\
~ & ~ \\
$\eta_2^{\pm}\rightarrow W^{\pm} \eta_1 \rightarrow  W^{\pm} \gamma \gamma$ & $0-0.22$ \\
$\eta_2^{\pm}\rightarrow W^{\pm} \eta_1 \rightarrow  W^{\pm} \gamma Z $ & $0-0.22$ \\
$\eta_2^{\pm}\rightarrow W^{\pm} \eta_1 \rightarrow  W^{\pm} Z Z $ & $0-0.18$ \\
~ & ~ \\
$\eta_1^{\pm}\rightarrow W^{\pm}V (V=\gamma, Z) $ & $\sim 1$ \\
\hline
\end{tabular}
\caption{Branching ratios for cascade decays of the charged pNGBs, when the parent pNGB scalar($\eta_5^{\pm \pm}$, $\eta_2^{\pm}$) decays to an on-shell gauge boson and another on-shell pNGB scalar, for the fermiophobic case. We have fixed $f=5$ TeV and all scalars have masses in the range $\sim 1.5-2$ TeV. Note that, being the lightest scalar, $\eta_1^{\pm}$ has $100$\% branching ratio to gauge bosons, occurring via off-shell pNGBs.}
\label{table:decay1}
\end{table}
\begin{table}[h!]
\centering
\begin{tabular}{|c|c|}
\hline
{\bf Casecade Decay} & {\bf Branching Ratio} \\
\hline
$\eta_5^{\pm \pm}\rightarrow  W^{\pm} \eta_1^{\pm} \rightarrow  W^{\pm}  W^{\pm} t t$ & $0.39-0.94$\\ 
~ & ~ \\
$\eta_2^{\pm}\rightarrow W^{\pm} \eta_1 \rightarrow  W^{\pm} t t $ & $0-0.3$ \\
~ & ~ \\
$\eta_1^{\pm}\rightarrow W^{\pm} t t$ & $\sim 1$\\
\hline
\end{tabular}
\caption{Branching ratios for cascade decays of the charged pNGBs, when the parent pNGB scalar($\eta_5^{\pm \pm}$, $\eta_2^{\pm}$) decays to an on-shell gauge boson and another on-shell pNGB scalar, for the fermiophilic case. We have fixed $f=5$ TeV and all scalars have masses in the range $\sim 1.5-2$ TeV. Note that, being the lightest scalar, $\eta_1^{\pm}$ has $100$\% branching ratio to $W^{\pm}tt$, occurring via off-shell pNGBs.}
\label{table:decay2}
\end{table}

\begin{figure}[h!]
\begin{center}
\scalebox{1.2}{
\includegraphics[width=6cm,height=5cm]{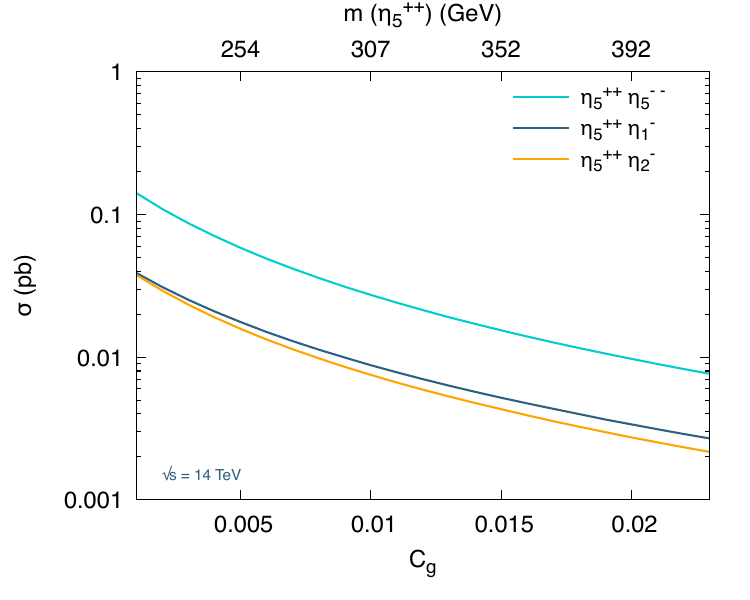}
\includegraphics[width=6cm,height=5cm]{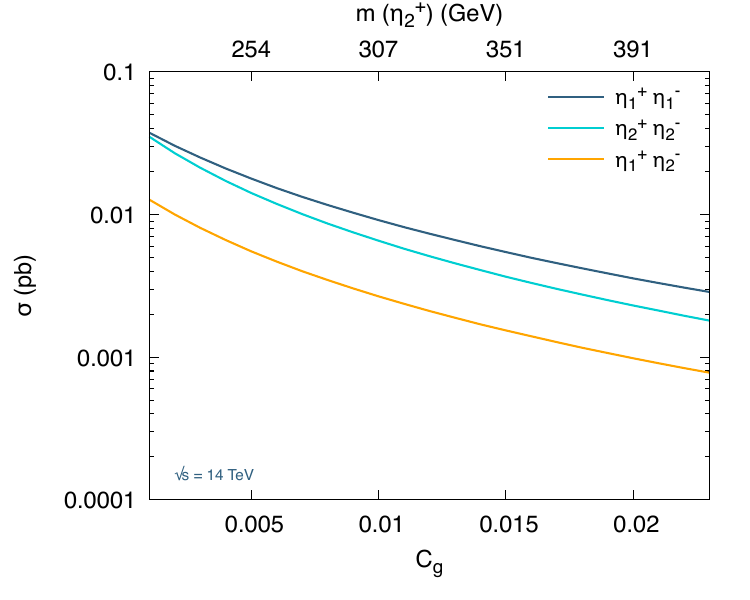}}
\caption{Production cross-sections of the charged scalars at the LHC $14$ TeV with $f=1$ TeV.}
\label{fig:lhc_14tev_f_1tev}
\end{center}
\end{figure}

In Fig.~\ref{fig:lhc_14tev_f_1tev}, we plot the production cross-section 
of the charged pNGB scalars at the $14$ TeV LHC. The scalar masses are represented as a function of $C_g$ while 
the value of \emph{f} is fixed at $1$ TeV. At this value of \emph{f}, the masses of the scalars are below $500$ GeV.
The decays that we are concerned about occur mostly at $f=5$ TeV and $C_g\gtrsim0.01$ \emph{i.e.,} at pNGB scalar masses above $1$ TeV.
For charged scalars in the TeV range, the cross-section at the LHC is estimated to be $\sim {\cal{O}}(1)$ fb or even smaller. 
Further, the signatures of interest contain multiple $W/Z$ bosons which lead to multi-lepton and/or multi-jet final states. When the $W/Z$ bosons decay leptonically, the resulting signals are suppressed by the small leptonic branching ratios. On the other hand, the hadronic decays avoid this suppression but suffer from substantial QCD backgrounds at the LHC.

\begin{figure}[h!]
\begin{center}
\scalebox{1.2}{
\includegraphics[width=6cm,height=5cm]{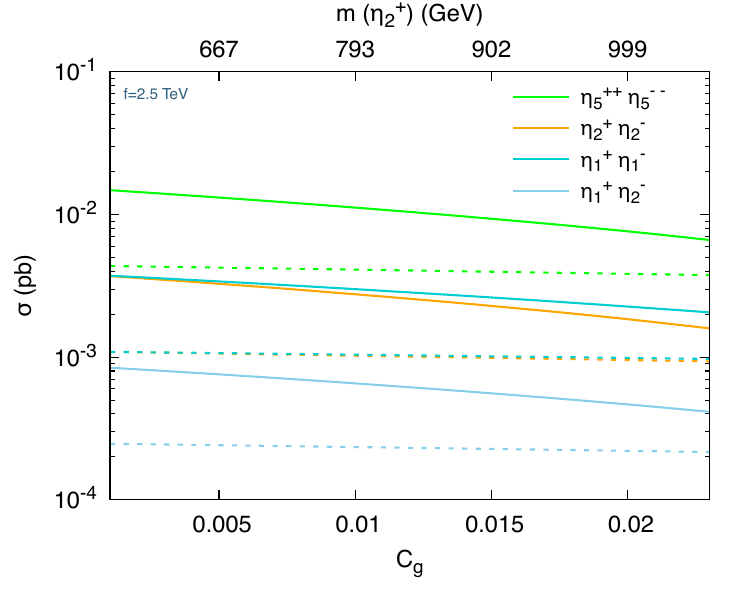}
\includegraphics[width=6cm,height=5cm]{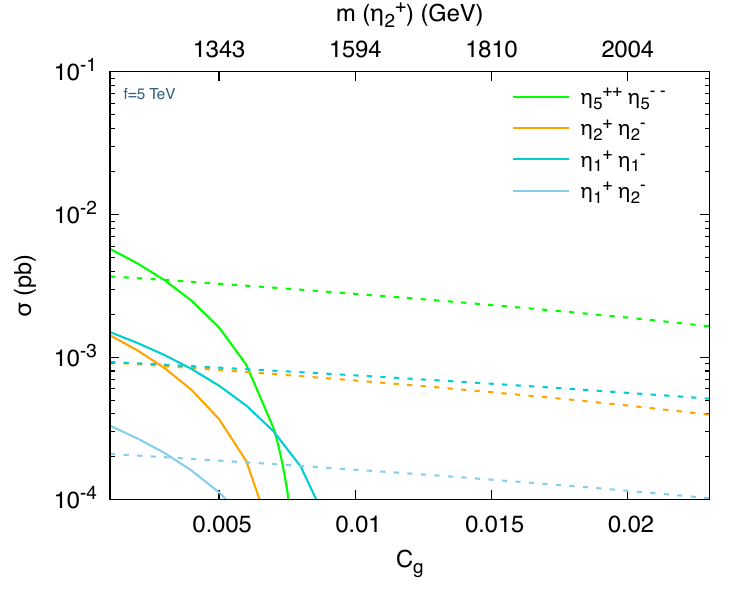}}
\caption{Production cross-sections of the pNGB scalars at $\mu^+\mu^-$ collider, at $f=2.5$ TeV (Left) and $f=5$ TeV (Right), for $\sqrt{s}=3~TeV$ (solid lines) and $\sqrt{s}=6~TeV$ (dashed lines).}
\label{fig:muon_collider}
\end{center}
\end{figure}

For a cleaner environment, lepton colliders promise to be more effective than the $pp$ collider. 
However, the future lepton colliders such as ILC and CLIC  ~\cite{Kumar:2024ghy,Lukic:2014mha,ILCInternationalDevelopmentTeam:2022izu} 
will have their achievable center-of-mass energies limited to the sub-TeV regime. Since we are interested in the $W/Z$ rich signatures of pNGB scalars with masses exceeding the TeV scale, the muon collider ~\cite{Han:2022edd,Delahaye:2019omf,InternationalMuonCollider:2024jyv,Accettura:2023ked} will be a great alternative.
Unlike the lepton colliders with electron beams, the energy loss due to the synchrotron radiation is suppressed in muon colliders because of the large mass of muon. Thus, achieving both high energy as well as high luminosity is possible.
We plot the production cross-section of the pNGB scalars at the $3$ TeV and $6$ TeV Muon collider in Fig.~\ref{fig:muon_collider}.
The pair production cross-section of the two singly charged scalars, $\eta_2^+ \eta_2^-$ and $\eta_1^+ \eta_1^-$, are different due to difference in their gauge couplings. The associated pair production cross-section, $\eta_1^+ \eta_2^-$, is comparatively smaller than the pair production cross-sections.
 
The pair production of charged scalars, followed by cascade decays, gives rise to a variety of collider signatures such as monolepton, dilepton, and trilepton signals along with multiple jets. 
As discussed earlier, we are interested in the decays arising in the regime $f=5$ TeV with scalar masses above $1.5$ TeV. Hence, we adopt this benchmark setup for the signal analysis and choose $C_g=0.015$ and $f=5$ TeV.
In the fermiophobic scenario, the two heavier charged pNGB scalars undergo two-body on-shell decays \emph{viz.,} 
$\eta_2^\pm \rightarrow W^\pm \eta_1$ and $\eta_5^{\pm\pm} \rightarrow W^\pm \eta_\pm$. Considering further the decays of $\eta_1$ and $\eta_1^\pm$ via their dominating modes, the possible signatures are,
\begin{enumerate}
\item $\mu^+ \mu^- \rightarrow \eta_2^+ \eta_2^-\rightarrow (W^+ \eta_1)(W^- \eta_1)\rightarrow (W^+ ZZ)(W^+ ZZ)\rightarrow 4f_j + (\ell^+ \ell^-)+ MET $
\item $\mu^+ \mu^- \rightarrow \eta_5^{++} \eta_5^{--}\rightarrow (W^+ \eta_1^+)(W^- \eta_1^-)\rightarrow (W^+ W^+Z)(W^- W^- Z)\rightarrow 2f_j + 2(\ell^+ \ell^-)+ MET $.
\end{enumerate}

These signals are rich in highly energetic $W$ and $Z$ bosons which can be identified as {\it fatjets} ($f_j$). 
In both cases, we consider the possibility when $W$ decays to the leptons and $Z$ decays to jets. Hence, $Z$ will be identified as a fatjet.
Up to the intermediate state consisting of all gauge bosons, the cross-section $\times$ BR in these channels are $\sim$ $0.02$ fb and $0.2$ fb, respectively, at the $6$ TeV muon collider.
If we consider the pair production of $\eta_1^+ \eta_1^-$, we obtain the final state, $W^+ZW^-Z$ that leads to dilepton $+ ~2/3 ~f_j$ signals.
The dominant backgrounds for these channels are $nV ~+$ jets, where $n$ is the number of $W/Z$ bosons. The leptons originate from the decay of the $W/Z$ bosons and the highly energetic QCD jets are identified as fatjets. However, $W/Z$ bosons in SM background processes, have lesser chance of being identified as fatjets as they have less energy. By imposing cuts on the fatjet mass and other jet-substructure variables, these backgrounds can be reduced. Events involving $nV ~+$ jets +$\gamma$ also contributes as a major background; however, the cross section is found to be typically an order of magnitude smaller than $nV ~+$ jets.

In addition, when we consider the alternate decay modes of $\eta_1$ and $\eta_1^{\pm}$, the following signals arise:
\begin{enumerate}
	\setcounter{enumi}{2}
\item $\mu^+ \mu^- \rightarrow \eta_2^+ \eta_2^-\rightarrow (W^+ \eta_1)(W^- \eta_1)\rightarrow (W^+ W^+W^-\gamma)(W^+ W^+W^-\gamma)\rightarrow 4f_j + (\ell^+ \ell^-)+ (\gamma\gamma)+ MET $
\item $\mu^+ \mu^- \rightarrow \eta_5^{++} \eta_5^{--}\rightarrow (W^+ \eta_1^+)(W^- \eta_1^-)\rightarrow (W^+ VVV)(W^- VVV)\rightarrow n f_j + m(\ell^+ \ell^-)+ MET  $
\end{enumerate}
where $V=W^\pm, \gamma, Z$. These signals are also rich in fatjets, arising from the decays of $W$ and/or $Z$. 
The resulting fatjet and lepton multiplicities in signal $4$ depend upon the number of $W$ and $Z$ bosons present in the decay chain and their subsequent decays, and are hence, denoted by $m$ and $n$, respectively.
Up to the intermediate state consisting of all gauge bosons, the cross-section $\times$ BR at the $6$ TeV muon collider are $0.002$ fb and $0.014$ fb, for channels 3 and 4, respectively. These are comparably smaller than for channels 1 and 2.

To analyze these fatjet-rich signals, we generate the signal samples for a muon collider operating at c.m. energy, $\sqrt{s}=6$ TeV using \textsc{MadGraph5\_aMC@NLO}~\cite{Alwall:2014hca}, importing the corresponding UFO model file. \textsc{Pythia \;8.3}~\cite{10.21468/SciPostPhysCodeb.8} is used for the hadronisation of quarks and gluons, while \textsc{Delphes}~\cite{deFavereau2014}, with the muon collider detector card, is used to simulate particle interactions in the detector.
As the signals are rich in fatjets, the signal sensitivity will strongly depend on the fatjet reconstruction efficiency of the detector.
\begin{figure}[h!]
\includegraphics[height=4.8cm,width=5.7cm]{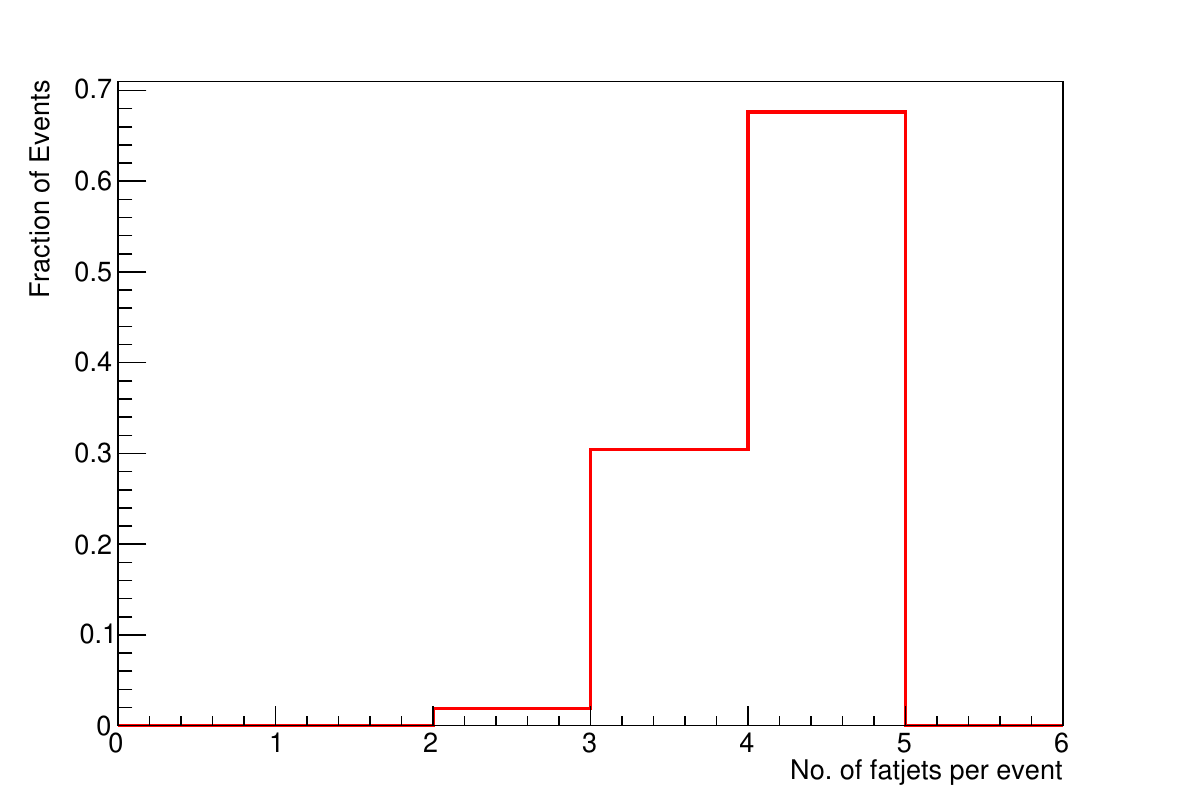}
\includegraphics[height=4.8cm,width=5.7cm]{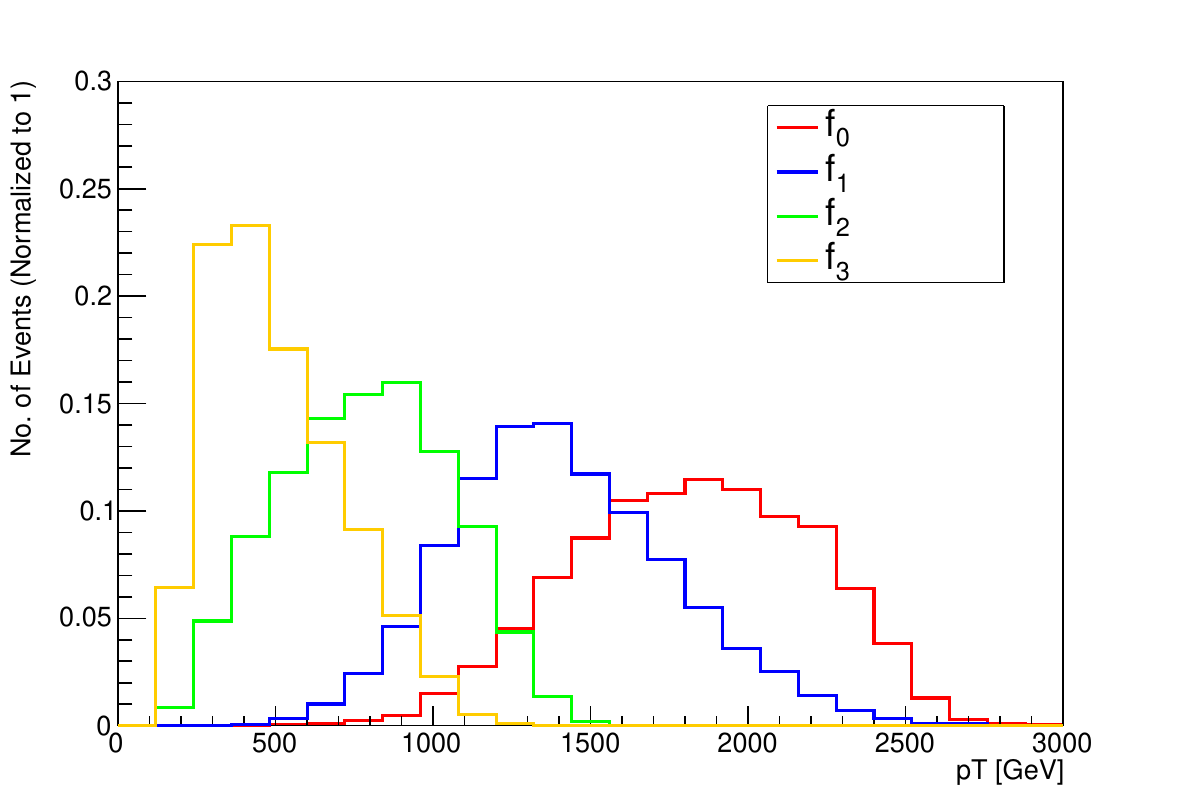}
\includegraphics[height=4.8cm,width=5.7cm]{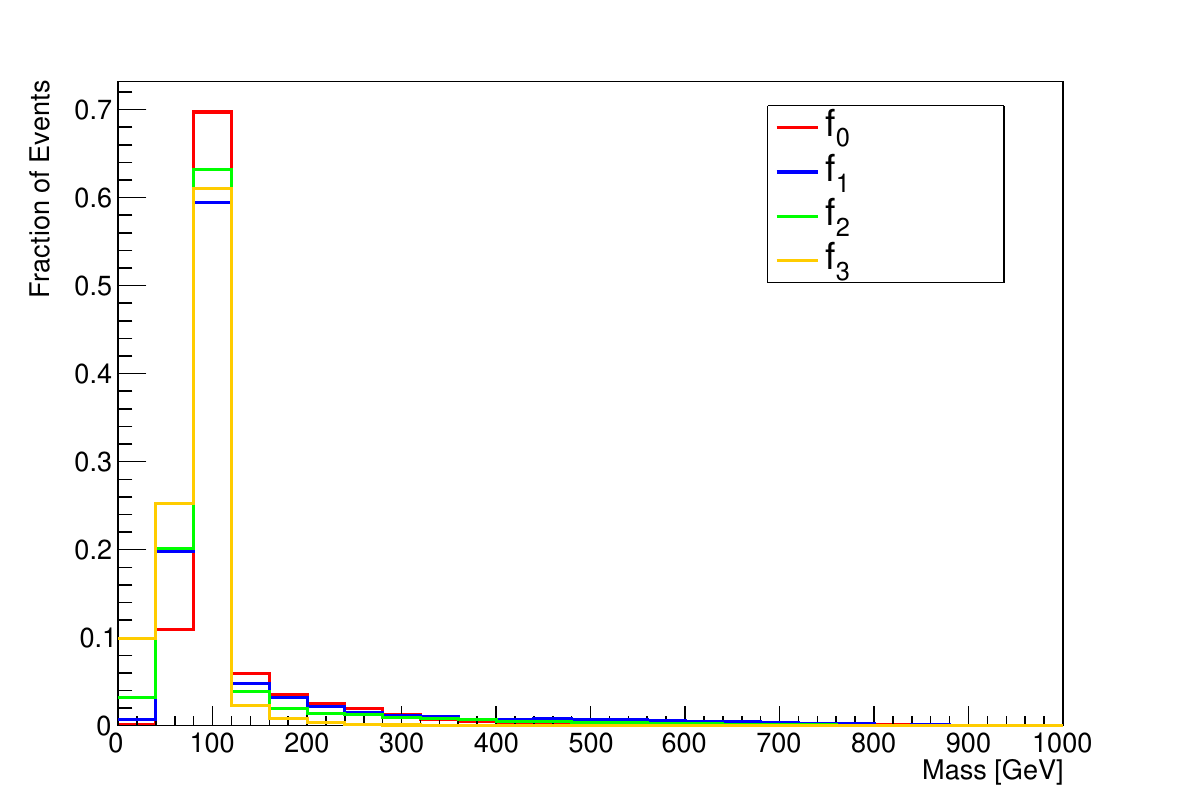} \\
\includegraphics[height=4.8cm,width=5.7cm]{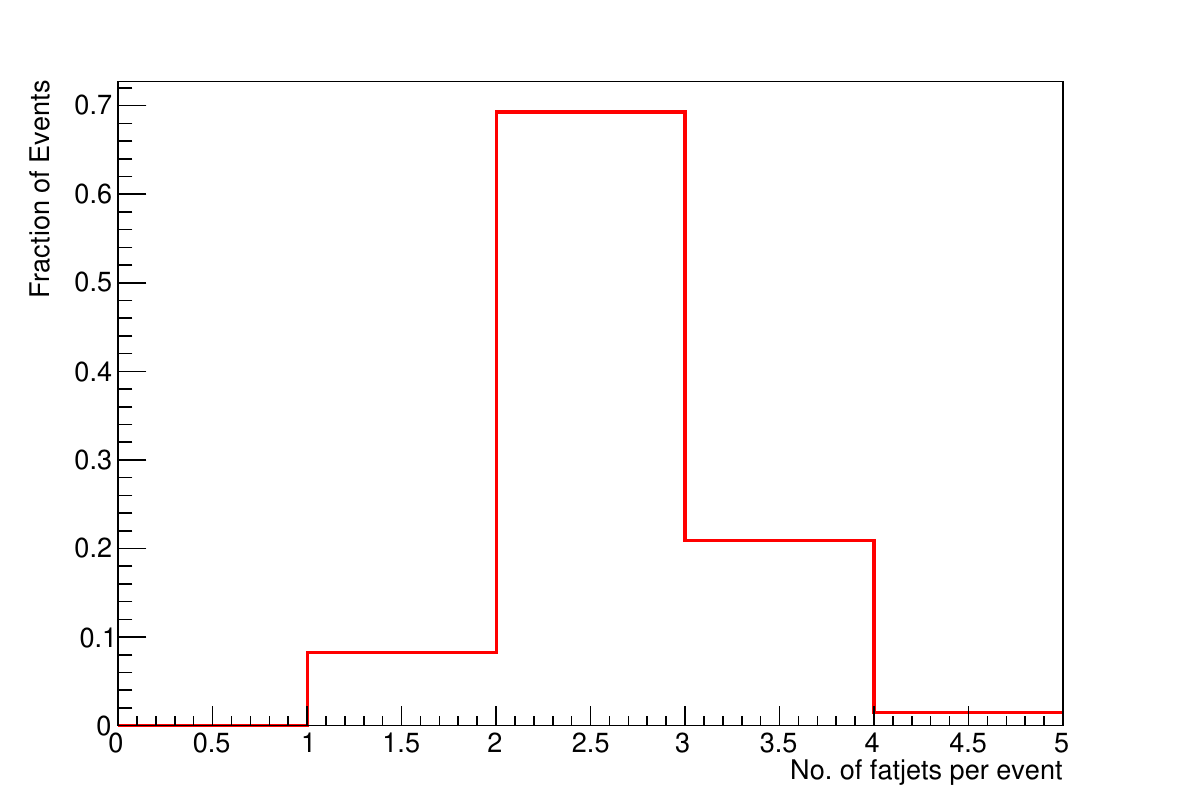}
\includegraphics[height=4.8cm,width=5.7cm]{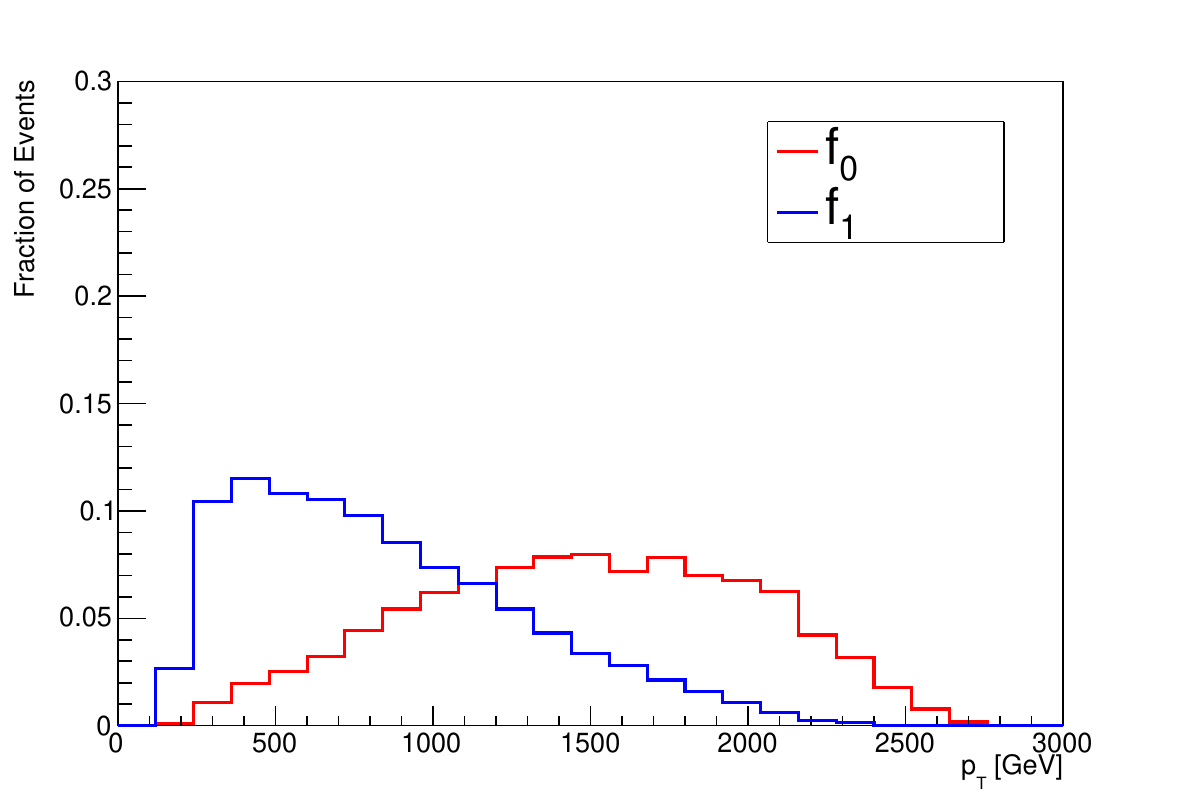}
\includegraphics[height=4.8cm,width=5.7cm]{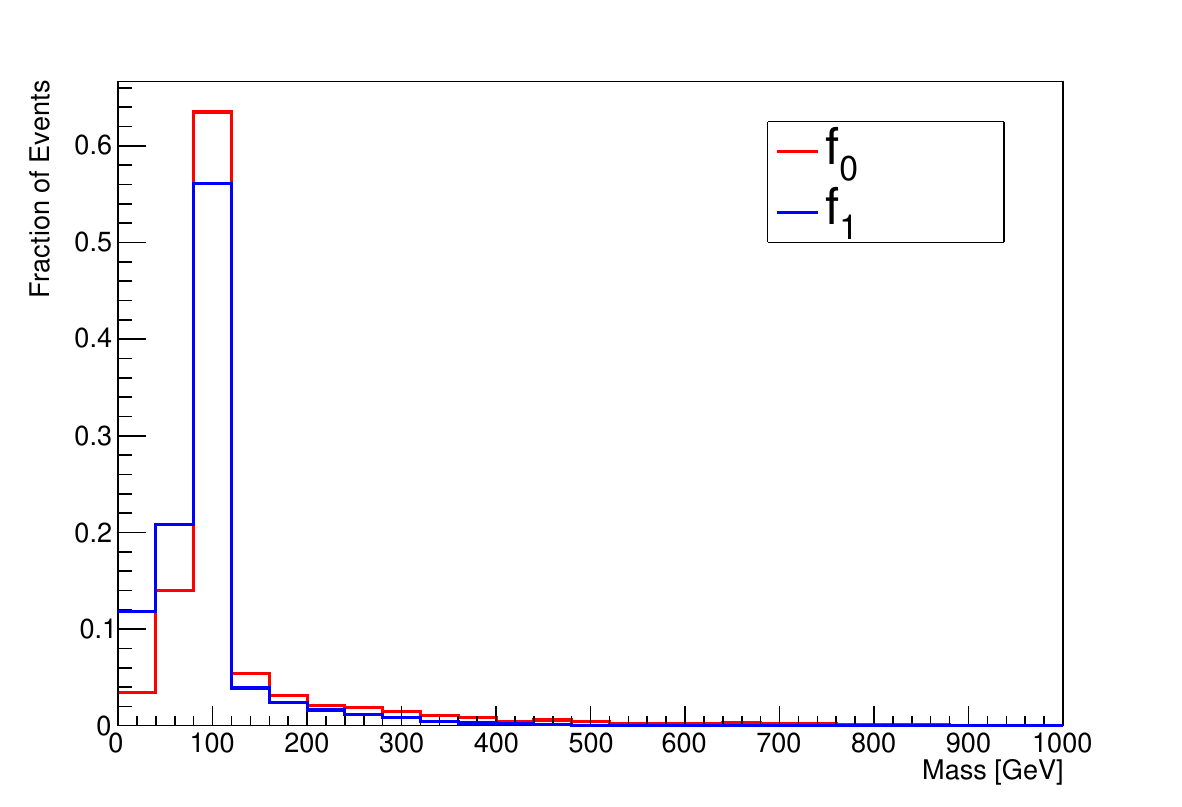}
    \caption{Distribution of kinematic variables for Signal 1 (Top) and Signal 2 (Bottom), at the $6$ TeV muon collider. In all plots, total number of events is normalized to 1.}
    \label{fig:fatjet}
\end{figure}

For studying the kinematical variables and their event distributions reliably, we apply initial cuts on the kinematical variables. We select leptons with $p_T(e,\mu)>10$ GeV, $|\eta|(e,\mu) <2.4$, $p_T(\tau)>20$GeV and $|\eta|(\tau)<2.3$. The jets
and the fatjets are reconstructed using the VLC (Valencia) jet algorithm \cite{Boronat:2014hva}. 
Jets are selected with jet radius of $0.5$, $p_T(j)>30$ GeV, $|\eta|(j) < 2.4$ and $\Delta R (\ell j)>0.4$ and fatjets are selected with jet radius $R=1.0$, $p_T(f_j)>200$ GeV and $|\eta(f_j)|<2.4$ and $\Delta R (\ell f_j)>0.4$.
In Fig.~\ref{fig:fatjet}, we plot the distributions of fatjet multiplicity (left), transverse momentum ($p_T$) (center) and invariant mass (right), for the signal 1 and 2. From Fig.~\ref{fig:fatjet} (top and bottom center), we can observe that the fatjet $p_T$ is very high and the mass of the fatjets are also correctly reconstructed.

We find that the correct number of fatjets are identified in almost $70$\% of the events. However, we have found that the number of events with fatjet strongly depends on two factors. (i) Fatjet Radius (R): If we choose different values of fatjet radius, say $R=0.8$ or $1.2$, 
the fatjet identificantion efficiency changes, and the number of events with fatjets are less or more. (ii) Jet Reconstruction Algorithm: 
If we use other algorithms in Delphes such as anti-kt \cite{Cacciari:2008gp} or N-subjettiness algorithm \cite{Thaler:2010tr}, the fatjet identification efficiency changes. 
Thus the signal efficiency is affected as the number of fatjets are very high in these signals. 
The fatjet masses are reconstructed successfully up to the third fatjet ($f_2$) for signal 1 (top right). Selection of jet pruning algorithm plays a major role in the fatjet mass reconstruction.
Also, Beam-induced backgrounds (BIB) are anticipated to be significant at a muon collider. 
A full treatment of BIB requires a dedicated simulation framework, optimized timing cuts, and detector-specific mitigation strategies. 
In addition, the cascade decays give rise to final states with a large number of particles,
in contrast to the final states targeted in conventional searches of the scalars. 
Given the small signal as well as background cross-sections in these channels, 
machine-learning techniques are expected to play a crucial role in efficient discrimination of signal from the background.
Overall, the signal efficiency of the fatjet rich signals are found out to be promising but it depends on several parameters such as jet algorithm, jet radius, jet pruning algorithm and other jet substructure variables, thereby requiring a dedicated study.

\section{Conclusion}
Key unanswered questions that remain after the discovery of Higgs are about the fundamental versus composite nature of Higgs, its unprotected mass from high-scale quantum corrections, and the absence of a dynamical electroweak symmetry breaking (EWSB) mechanism in the SM. Composite Higgs Models address these issues by introducing a TeV-scale strongly interacting sector where the Higgs emerges  as a pseudo-Nambu-Goldstone boson (pNGB) from spontaneous symmetry breaking. In the non-minimal composite model based on $SU(5)/SO(5)$ framework, resembling the Georgi-Machacek model at low 
energies, features a rich scalar sector. The fermiophilic and fermiophobic decay channels for pNGBs are influenced by fermion embedding choices in $SU(5)/SO(5)$ representations. 

Recent studies have examined the fermiophilic and fermiophobic scenarios, analyzing collider signatures and the LHC constraints under simplified assumptions with fixed compositeness 
scale \emph{f} and scalar masses. However, the compositeness scale \emph{f} (varied from $1–5$ TeV) is intrinsically linked to fine-tuning, which depends on the choice of representation 
for the fermions. In this work, we have shown that the masses and mixing of pNGB scalars, as functions of \emph{f} and gauge-loop parameter $C_g$, have a key influence on the branching ratios of 
the pNGB scalars. Unlike the prior works, this study incorporates mixing between gauge and mass eigenstates, deriving couplings in terms of the mixing angles, which influences the scalar phenomenology.

In this paper, we have studied the fermiophilic and fermiophobic scenarios for the full scalar spectrum, where, the scalar masses fall in the range $\sim (200-2000)$ GeV. We find that the decay patterns of the scalars differ not only across the two scenarios but also within each scenario, as a function of the masswithout
The on-shell and off-shell decay modes are highly dependent on the mass differences among the pNGB scalars, 
which is calculated explicitly in terms of the model parameters $f$, $C_g$ and mixing angles. 
Further, we show that the branching ratios of the two singly charged scalars, $\eta_1^{\pm}$ and $\eta_2^{\pm}$ are very different. 
In the fermiophobic scenario, at small masses, both the scalars decay to the gauge bosons but at large masses, their decay patterns change. 
The heavier $\eta_2^{\pm}$ decays to another on-shell pNGB scalar, accompanied by an off-shell or on-shell gauge boson. On the other hand, the scalar, $\eta_1^{\pm}$, due to its small mass splitting with the other pNGBs, cannot decay to an on-shell pNGB scalar. Instead, it prefers decaying to two gauge bosons. 
In the fermiophilic scenario, $\eta_2^{\pm}$ decays dominantly to a top and a bottom quark, at small and intermediate masses and to an on-shell pNGB and a $W$ boson, at large masses. 
The scalar, $\eta_1^{\pm}$, instead, prefers decaying to a top and a bottom quark at small masses, to $Z\bar{t}b (Z\bar{b}t)$ at intermediate masses and $W^{\pm}\bar{t}t$ at large masses. 

For the collider study, we focus on the decay channels where the parent pNGB scalar decays to a lighter pNGB scalar and an on-shell gauge boson. The daughter pNGB scalar decays further to two gauge bosons leading to a cascade decay. These processes occur in the high-mass regime, with final states comprising multiple leptons and highly bosted jets. However if these jets are coming from the gauge bosons, then there is a greater probability of identifing $W/Z$ bosons as fatjet. A muon collider is therefore, particularly well suited for probing such signatures at large energy. We estimate the cross-section $\times$ BR in the multilepton+fatjet channels when the charged pNGB scalars are pair produced at $6$ TeV muon collider. We discuss the backgrounds for these channels and the kinematical distributions which are crutial for the  identification of the fatjets. 
Given that the signal involves multiple fatjets, it is crucial to implement a complete treatment using several fatjet reconstruction methods. A detailed examination of the beam-induced background is also necessary. Moreover, the kinematics feature cascade decays involving two distinct BSM particles, necessitating sophisticated approaches like machine learning techniques. Therefore, a thorough analysis is beyond the capacity of this study and will be pursued in subsequent work.

\section*{Acknowledgments}
N.K expresses gratitude to A.M. Anirudhan for his contributions during the initial stages of this project, where he served as the project associate (PA-1) under SRG/2022/000363. 
N.K. would like to thank Department of Science and Technology and Anusandhan National Research Foundation
Government of India for the support from the Startup Research Grant, grant agreement number SRG/2022/000363 and Core Research Grant with grant agreement number CRG/2022/004120.